\documentclass[lettersize,journal]{IEEEtran}

\usepackage[numbers,sort&compress]{natbib}
\usepackage{url}
\usepackage{verbatim}
\usepackage{amsthm,amsmath,amsfonts,amssymb}
\usepackage{mathrsfs}
\usepackage{graphicx}
\usepackage{stfloats}
\usepackage{caption}
\usepackage{array}
\usepackage{booktabs}
\usepackage{multirow}
\usepackage[ruled, vlined, linesnumbered]{algorithm2e}
\usepackage{setspace}
\usepackage{pgfplots}
\usepackage{etoolbox}
\usepackage{pifont}
\apptocmd{\thebibliography}{\small}{}{}

\SetAlgoCaptionSeparator{.}

\SetAlgoLined
\SetEndCharOfAlgoLine{}
\SetKwProg{Fn}{}{}{}

\begin{document}
\title{A Distributed Scalable Cross-chain State Channel Scheme Based on Recursive State Synchronization}
	
\author{Xinyu Liang, Ruiying Du, Jing Chen, Yu Zhang, Meng Jia, Shuangxi Cao, Yufeng Wei, Shixiong Yao
	
	\thanks{This research of Wuhan University was supported in part by the open project of Satellite Internet Key Laboratory in 2022 (Project 6: Research on Distributed Trusted Communication Networking Technology of Blockchain for Satellite Internet). (\emph{Corresponding author: Ruiying Du})}
	\thanks{Xinyu Liang, Ruiying Du, Jing Chen and Meng Jia are with the School of Cyber Science and Engineering, Wuhan University, Wuhan 430072, China (e-mail: liangxinyu@whu.edu.cn; duraying@whu.edu.cn; chenjing@whu.edu.cn; jiameng@whu.edu.cn)}
	\thanks{Yu Zhang is with the Beijing Infosec Technologies Co., Ltd, Beijing 100089, China (e-mail: zy168612@163.com)}
	\thanks{Shuangxi Cao and Yufeng Wei are with the China Satellite Network Innovation Co., Ltd, Beijing 100000, China (e-mail: caosxhcscn@163.com; weiyufeng59@126.com)}
	\thanks{Shixiong Yao is with the School of Computer Science, Central China Normal University, Wuhan 430079, China (e-mail: yaosx@ccnu.edu.cn)}
}

\maketitle

\begin{abstract}
As cross-chain technology continues to advance, the scale of cross-chain transactions is experiencing significant expansion. To improve scalability, researchers have turned to the study of cross-chain state channels. However, most of the existing schemes rely on trusted parties to support channel operations. To address this issue, we present Interpipe: a distributed cross-chain state channel scheme. Specifically, we propose a real-time cross-chain synchronization scheme to ensure consistent operations between two blockchains to a cross-chain state channel. Moreover, we propose a batch transaction proof scheme based on recursive SNARK to meet the cross-chain verification needs of large-scale users. Based on the above designs, Interpipe offers protocols for opening, updating, closing, and disputing operations to cross-chain state channels. Security analysis shows that Interpipe has consistency and resistance, and experimental results demonstrate that a cross-chain state channel can be nearly as efficient as an existing intra-chain state channel.
\end{abstract}


\section{Introduction} \label{Section_Introduction}
Blockchain \cite{Paper_Bitcoin} offers a secure and transparent way to record and verify transactions without the need for a central authority. In recent years, distributed trust systems centered around blockchains have been forming at an unprecedented pace. The application areas \cite{Paper_BlockchainApplication} of blockchain have expanded from the traditional cryptocurrency domain \cite{Paper_Bitcoin} \cite{Paper_Ethereum} to various fields such as healthcare \cite{Paper_BlockchainInHealthcare}, supply chain \cite{Paper_BlockchainInSupplyChain}, and cloud service \cite{Paper_BlockchainInCloudService}. However, while enriching the blockchain ecosystem, these blockchains are isolated from each other, hindering the flow of information and value. To solve the isolation problem, cross-chain technology \cite{Paper_VitalikButerin} enables different blockchain networks to communicate and interact with each other, which is invaluable for connecting the decentralized Web 3.0.

The existing cross-chain schemes can be classified into two categories, which are non-relay-based scheme and relay-based scheme. The non-relay-based scheme appeared in the early stage of cross-chain technology. It relies on external components to achieve cross-chain operations, such as decentralized exchanges \cite{Paper_DecentralizedExchange} based on notaries, and atomic cross-chain swaps \cite{Paper_AtomicSwap} based on hashed time-lock contract \cite{Paper_MAD-HTLC}. However, non-relay-based schemes do not have the transfer of state information between blockchains, therefore limiting their functionality to relatively simple cross-chain operations. In a relay-based scheme \cite{Paper_TwoWayPeg} \cite{Paper_BTCRelay} \cite{Paper_XClaim} \cite{Paper_zkbridge} \cite{Paper_Polkadot} \cite{Paper_Cosmos} \cite{Paper_XPull}, blockchains transfer their state information to each other by using a cluster of relay nodes. Based on the state information, a party of a blockchain can directly verify the transactions in another blockchain to support more complex cross-chain operations, leading to the emergence of cross-chain platforms such as Polkadot \cite{Paper_Polkadot} and Cosmos \cite{Paper_Cosmos}. With the number of users on the platform increasing, the daily cross-chain transactions have extended to a considerable scale, which will eventually exceed the blockchain throughput limit \cite{Paper_HuobiResearch}, resulting in a scalability issue.

In previous intra-chain scenarios, the scalability issue can be solved by state channel \cite{Paper_LightningNetwork} \cite{Paper_PaymentChannelPrivacy} \cite{Paper_PaymentChannelAtomicity} \cite{Paper_PaymentChannelThora} \cite{Paper_PaymentChannelSleepy} \cite{Paper_PaymentChannelScheduling} \cite{Paper_Perun} \cite{Paper_Multi-partyVirtualStateChannels} \cite{Paper_Bitcoin-compatibleVirtualChannels}. In the most recent years, concerning the idea of state channel, researchers have begun to study the potential of \emph{cross-chain state channel} between two blockchains, and move most cross-chain transactions into the channel for executions to share the blockchain throughput pressure. As far as we know, there are currently two papers \cite{Paper_Cross-chainVirtualPaymentChannels} \cite{Paper_CrossChannel} that provide detailed designs. Specifically, Jia et al. propose cross-chain virtual payment channel \cite{Paper_Cross-chainVirtualPaymentChannels} to achieve off-chain interactions between Ethereum and Bitcoin. Guo et al. propose Cross-Channel \cite{Paper_CrossChannel} to support cross-chain operations in both synchronous and asynchronous networks. However, these schemes rely on trusted parties to support channel operations, which are vulnerable in a distributed environment. To achieve a distributed cross-chain state channel, we are still facing two critical challenges.

\begin{enumerate}
	\item \textbf{Consistent Operation:} Two blockchains must synchronize their state information to ensure consistent operations within a cross-chain state channel. However, in most existing schemes, cross-chain synchronizations are non-real-time, leaving room for third parties or intermediaries to arbitrarily delay or interrupt the synchronization process. This can lead to situations where two blockchains have different operations within one cross-chain state channel, posing a security threat.
	
	\item \textbf{Scalable Verification:} Operations within cross-chain state channels require blockchain nodes to efficiently verify transactions on another blockchain. However, most of the existing schemes focus on verifying specific individual transactions. When the scale of the transactions becomes large, the cross-chain verifications will incur significant costs, which cannot satisfy the growing demands of users.
\end{enumerate}

In this paper, we present Interpipe: a distributed cross-chain state channel scheme. To ensure consistent operations, we introduce a real-time cross-chain synchronization scheme. Specifically, our approach involves the adoption of a state pulling strategy to retrieve the latest state from one blockchain and generate the corresponding state proof to be recorded in another blockchain. This process operates recursively in the background as the blockchain expands, a method we refer to as recursive state synchronization. As a result, two blockchains can synchronize their real-time state proofs with each other. Based on state synchronization, we achieve transaction synchronization. It enables the recording of one transaction $ct$ into two blockchains, while maintaining their consistency based on the real-time state proofs of each other. In the subsequent discussion, $ct$ is referred to as a cross-chain transaction. Subsequently, cross-chain transactions can be published on both blockchains to facilitate the operations of opening, updating, closing, and disputing within a cross-chain state channel.

To facilitate scalable verification, we propose a batch transaction proof scheme based on recursive SNARK. Specifically, within the system of blockchain $\mathbf{P}_i$, a prover aggregates every cross-chain transaction in $\mathbf{P}_i$ into a one-way accumulator and binds this accumulator with the current state proof of $\mathbf{P}_i$ using zk-SNARK. Notably, we leverage the recursively generated data structure of the blockchain to implement recursive SNARK, thereby reducing the computational cost of generating the state proof. On the other hand, within the blockchain system $\mathbf{P}_j$, a verifier first verifies the correctness of the accumulator based on the state proof of $\mathbf{P}_i$. Subsequently, each cross-chain transaction in $\mathbf{P}_i$ can be efficiently verified based on the accumulator. Therefore, the batch proof system caters to the verification needs of any cross-chain transaction, meeting the large-scale requirements of users.

In general, we have the following contributions.
\begin{enumerate}
	\item Interpipe represents the first distributed cross-chain state channel scheme. To the best of our knowledge, existing cross-chain state channel schemes rely on trusted parties to support channel operations, which introduces security vulnerabilities.
	
	\item To achieve consistent operations, we propose a real-time cross-chain synchronization scheme. This enables us to record one cross-chain transaction into two blockchains and ensure their consistency. Subsequently, we carry out operations of opening, updating, closing, and disputing within a cross-chain state channel.
	
	\item To achieve scalable verification, we propose a batch transaction proof scheme based on recursive SNARK. This approach allows a verifier within a blockchain system to efficiently verify any cross-chain transaction in another blockchain.
	
	\item We conduct a security analysis of Interpipe. Additionally, we implement a proof-of-concept prototype and evaluate the performance of Interpipe.
\end{enumerate}

This paper is organized as follows. In Section \ref{Section_RelatedWork}, we review the existing works related to cross-chain scheme, state channel, and transaction verification. We describe the building blocks of our scheme in Section \ref{Section_Preliminary}, and give a description about the system model, threat model, and design goals in Section \ref{Section_ProblemStatement}. We illustrate the batch transaction proof in Section \ref{Section_BatchTransactionProof}, and describe the details of Interpipe in Section \ref{Section_Interpipe}. We analyze the security properties of our scheme in Section \ref{Section_SecurityAnalysis}, and describe the implementation of our prototype in Section \ref{Section_ExperimentAndEvaluation}. Finally, we conclude in Section \ref{Section_Conclusion}.

\section{Related Work} \label{Section_RelatedWork}
\subsection{Cross-chain Scheme} \label{Subsection_Cross-chainSchemes}
\begin{figure}[tbp]
	\centering
	\includegraphics[width=1\linewidth]{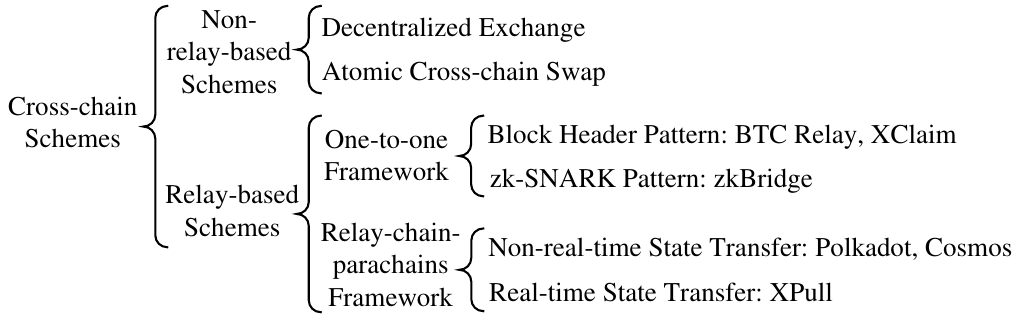}
	\caption{Categories of cross-chain schemes}
	\label{Figure_CategoriesOfCross-chainSchemes}
\end{figure}

As illustrated in Fig. \ref{Figure_CategoriesOfCross-chainSchemes}, the existing cross-chain schemes can be classified into two categories: non-relay-based schemes and relay-based schemes. In a non-relay-based scheme, two blockchains do not transfer their state information to each other but rely on external components to achieve cross-chain operations. For example, decentralized exchanges \cite{Paper_DecentralizedExchange} use a notary committee as a third party to exchange users' tokens in one blockchain for tokens in another blockchain. Moreover, atomic cross-chain swaps \cite{Paper_AtomicSwap} are achieved by using hashed time-lock contract \cite{Paper_MAD-HTLC}. However, as a non-relay-based scheme does not transfer blockchain state information, the scheme can only support simple cross-chain operations with a low security guarantee.

In a relay-based scheme, two blockchains transfer their state information to each other by using a cluster of relay nodes. The relay-based scheme was first illustrated by Back et al. \cite{Paper_TwoWayPeg} in 2014. BTC Relay \cite{Paper_BTCRelay} is a representative implementation. It sends the Bitcoin block headers to the Ethereum smart contract, achieving cross-chain verification of Bitcoin transactions. Based on BTC Relay, Alexei et al. \cite{Paper_XClaim} propose XClaim. It achieves trustless cross-chain exchanges using cryptocurrency-backed assets and employs collateralization and punishments to enforce the correct behavior of participants. To further improve efficiency, Xie et al. \cite{Paper_zkbridge} propose zkBridge. It introduces zk-SNARK to generate the state proof of blockchain. By this act, the state information can be compressed into a small proof to be transferred between blockchains, reducing the overhead of transmission and storage.

The early relay-based schemes mainly focus on the interaction between two blockchains, which is also called one-to-one framework. If a blockchain intends to interact with $n$ blockchains, it has to establish $n$ cross-chain connections, which is inefficient. To solve this problem, the relay nodes cluster begins to establish connections with multiple blockchains, called \textit{parachains} (parallel blockchains). Besides, the relay nodes maintain a blockchain, called \textit{relay chain}, by themselves to record the state information from every parachain. At the same time, relay chain is published to every parachain system. Subsequently, a parachain can obtain the states of other parachains by only accessing relay chain. Based on this relay-chain-parachain framework, a parachain only needs to establish one cross-chain connection with relay chain to interact with $n$ parachains that also connect with relay chain, to improve efficiency.

The existing cross-chain platforms such as Polkadot \cite{Paper_Polkadot} and Cosmos \cite{Paper_Cosmos} are constructed following the relay-chain-parachain framework. As mentioned in Section \ref{Section_Introduction}, these platforms are required to support the cross-chain needs of large-scale users with numerous cross-chain transactions per day. However, in previous cross-chain platforms, the cross-chain synchronizations between blockchains are non-real-time. It results that a blockchain cannot learn the latest states of another blockchain on the platform within a certain time limit, as the blockchain features a dynamically growing data structure with continuously updating states. To solve the problem, Liang et al. \cite{Paper_XPull} propose a cross-chain state pulling scheme, called XPull. The relay nodes cluster will periodically pull the latest state information from parachains, and forward the state information to other parachain systems, which achieves real-time state transfer to ensure timeliness.

\subsection{State Channel} \label{Subsection_StateChannel}
\begin{figure}[tbp]
	\centering
	\includegraphics[width=0.7\linewidth]{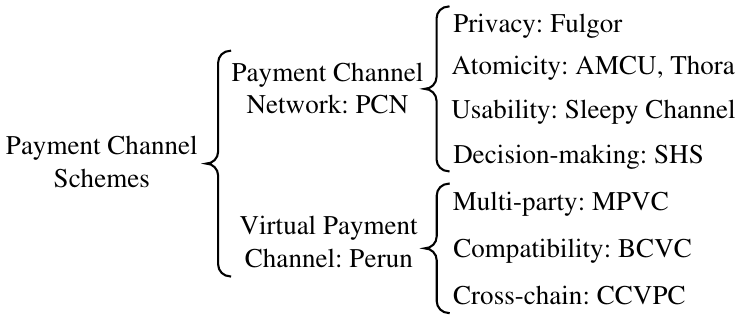}
	\caption{Categories of payment channel}
	\label{Figure_CategoriesOfPaymentChannel}
\end{figure}

In 2016, Joseph et al. proposed payment channel \cite{Paper_LightningNetwork} to improve the scalability of Bitcoin. In the following works, the payment channel is beginning to develop in two directions, as shown in Fig. \ref{Figure_CategoriesOfPaymentChannel}. The first direction is payment channel network \cite{Paper_LightningNetwork}, noted PCN. It enables the delivery of assets between two users $U_1$ and $U_n$ by using a path of payment channels in PCN. In the following works, researchers have further improvements to payment channel network from various perspectives. For example, Giulio et al. \cite{Paper_PaymentChannelPrivacy} propose Fulgor to ensure privacy by using multi-hop hashed time-lock contract. Christoph et al. \cite{Paper_PaymentChannelAtomicity} propose the protocol AMCU for atomic multi-channel updates by jointly creating a multiple-input-multiple-output (MIMO) transaction. Lukas et al. \cite{Paper_PaymentChannelThora} further propose Thora to refine atomicity, in which Thora is compatible with a number of cryptocurrencies having arbitrary payment channel topologies. Lukas et al. \cite{Paper_PaymentChannelSleepy} propose Sleepy Channel, which does not require either of the channel users to be persistently online. Papadis et al. \cite{Paper_PaymentChannelScheduling} propose single-hop scheduling (SHS), which provides a decision-making scheme for the users in payment channel network to maximize channel throughput.

The second direction is virtual payment channel \cite{Paper_Perun}. Based on the current two channels, noted as $(U_1, U_2)$ and $(U_2, U_3)$, the users $U_1$ and $U_3$ can establish a virtual payment channel $(U_1, U_3)$ to have direct interaction. Next, virtual payment channel technology has derived the design of multi-party virtual channel (MPVC) \cite{Paper_Multi-partyVirtualStateChannels}. It means that more than two users can interact within a channel, supporting more complex off-chain operations. Furthermore, Lukas et al. propose bitcoin-compatible virtual channel (BCVC) \cite{Paper_Bitcoin-compatibleVirtualChannels} to improve the compatibility. In the most recent work, Jia et al. propose a scheme (CCVPC) \cite{Paper_Cross-chainVirtualPaymentChannels} to achieve cross-chain virtual payment channel $(U_1, U_3)$, based on $(U_1, U_2)$ and $(U_2, U_3)$ in different blockchain systems. However, this design cannot resist the conspiracy attack of $U_1$, $U_2$, and $U_3$, posing a security threat.

Moreover, with the extension of blockchain application scenarios from cryptocurrencies to other fields, the operations of transferring tokens within a channel have been generalized into a channel's state change \cite{Paper_GeneralStateChannelNetworks}, in which the previous ``payment channel" has evolved into ``state channel". Nowadays, state channels have been applied in various fields, such as log audit \cite{Paper_DELIA}, data sharing \cite{Paper_DataSharing}, and video streaming \cite{Paper_VideoStreaming}.

\subsection{Transaction Verification} \label{Subsection_TransactionVerification}
The transaction verification schemes of blockchain can be classified into two categories. The first category is based on the well-known Merkle tree \cite{Paper_MerkleTree}, applied by Bitcoin. Subsequently, researchers improve the Merkle tree and propose Merkle Patricia tree, applied by Ethereum \cite{Paper_Ethereum}. It introduces prefix tree to enhance the efficiency of data storage and retrieval.

The second category is based on one-way accumulator. In 1993, Benaloh et al. \cite{Paper_One-wayAccumulator} first constructed the accumulator based on the one-way hash function which satisfies a quasi-commutative property. An accumulator allows a prover to hash the transactions in a blockchain into one short value, which supports efficient cross-chain transfer. Niko et al. \cite{Paper_CollisionFreeAccumulator} further generalize the definition of accumulators and construct a collision-free subtype. Jan et al. \cite{Paper_DynamicAccumulators} propose dynamic accumulator that allows a prover to dynamically add and delete an element in the accumulator. In recent works, Boneh et al. \cite{Paper_BatchingTechniquesForAccumulators} propose batching techniques for cryptographic accumulators, which achieve membership proof and non-membership proof of transactions in blockchain. As the accumulators are well compatible with zk-SNARK, the existing anonymous cryptocurrency schemes, such as Zcash \cite{Paper_Zerocoin} \cite{Paper_Zerocash}, adopt accumulators for transaction retrieval.

\section{Preliminary} \label{Section_Preliminary}
\subsection{Notation}
To facilitate the understanding, we summarize the main notations in this paper in TABLE \ref{Table_Notations}.

\begin{table}[h]
	\centering
	\footnotesize
	\caption{Notations}
	\label{Table_Notations}
	\begin{tabular}{c|l}
		\toprule
		\textbf{Notation}             &          \multicolumn{1}{c}{\textbf{Meaning}}                   \\
		\midrule
		$\mathbf{R}$                  & relay chain                                                     \\
		$\mathbf{P}_i$                & $i$-th parachain                                                \\
		$n$                           & number of parachains on a cross-chain platform                  \\
		$\mathrm{N}^R$                & set of nodes which maintain relay chain                         \\
		$\mathrm{N}^P_i$              & set of nodes which maintain the $i$-th parachain                \\
		$\mathrm{N}^{R \sim}_i$       & subset of $\mathrm{N}^R$ to interact with $\mathrm{N}^P_i$      \\
		$\pi^i$                       & zero knowledge proof of the state of $\mathbf{P}_i$             \\
		$A^i$                         & accumulator for every cross-chain transaction in $\mathbf{P}_i$ \\
		$ct$                          & cross-chain transaction                                         \\
		$w^{ct, i}$                   & membership witness of $ct$ in $\mathbf{P}_i$                    \\
		\bottomrule
	\end{tabular}
\end{table}

\subsection{Cross-chain State Transfer} \label{Subsection_Cross-chainStateTransfer}
The relay-chain-parachain framework is widely used in cross-chain platforms \cite{Paper_Polkadot} \cite{Paper_Cosmos} to meet the cross-chain needs of multiple blockchains.

There are $n$ parachains $\{\mathbf{P}_1,\dots,\mathbf{P}_n\}$ and one relay chain $\mathbf{R}$ in the framework. Each parachain $\mathbf{P}_i$ $(i \in [n])$ is maintained by a set of parachain nodes $\mathrm{N}^P_i$ and the relay chain $\mathbf{R}$ is maintained by a set of relay nodes $\mathrm{N}^R$. Moreover, $\mathrm{N}^R$ are divided into $n$ relay node groups $\{\mathrm{N}^{R \sim}_1,\dots,\mathrm{N}^{R \sim}_n\}$ and $ \forall \{i, j\} \subseteq \{1, \dots, n\} \wedge i \neq j, \mathrm{N}^{R \sim}_i \cap \mathrm{N}^{R \sim}_j = \emptyset$. The relay node group $\mathrm{N}^{R \sim}_i$ establish network connections with $\mathrm{N}^P_i$ to start the cross-chain state transfer, which includes two processes of state reception and state forwarding.

In the \emph{state reception} process, the relay node groups $\mathrm{N}^{R \sim}_i$ $(i \in [n])$ receive the state information of $\mathbf{P}_i$ from $\mathrm{N}^P_i$, and record the information into $\mathbf{R}$ to be published to every entity on the cross-chain platform. In the \emph{state forwarding} process, $\mathrm{N}^P_j$ $(j \in [n], i \neq j)$ extract the state information of $\mathbf{P}_i$ from $\mathbf{R}$, and record the information into $\mathbf{P}_j$. Therefore, by only accessing $\mathbf{P}_j$, any entity in $\mathbf{P}_j$ system can obtain the state information of $\mathbf{P}_i$, supporting cross-chain operations.

Furthermore, to ensure the timeliness of $\mathbf{P}_i$ state information in $\mathbf{R}$, the relay node group $\mathrm{N}^{R \sim}_i$ adopts a \emph{state pulling} strategy \cite{Paper_XPull}. $\mathrm{N}^{R \sim}_i$ periodically send the state pulling instructions to $\mathrm{N}^P_i$ to pull the latest state information of $\mathbf{P}_i$, and subsequently record the state information into $\mathbf{R}$. Moreover, when an adversary has corrupted $\mathrm{N}^{R \sim}_i$ to interrupt the state pulling, a new relay node group $\mathrm{N}^{*R \sim}_i$ will be randomly selected from $\mathrm{N}^R$ based on distributed randomness \cite{Paper_Randpiper} generated by $\mathrm{N}^R$. $\mathrm{N}^{*R \sim}_i$ will replace $\mathrm{N}^{R \sim}_i$ to continue the state pulling, ensuring $\mathbf{P}_j$ to record the real-time state information of $\mathbf{P}_i$.

\subsection{One-way Accumulator} \label{Subsection_One-wayAccumulator}
An accumulator \cite{Paper_One-wayAccumulator} \cite{Paper_CollisionFreeAccumulator} \cite{Paper_BatchingTechniquesForAccumulators} enables one to encode a set into a short digest and prove that an element is in the set. Let $D$ be the domain of an accumulator. An accumulator $\mathcal{ACC}=(\mathsf{Setup},\mathsf{Commit},\mathsf{Add},\mathsf{CreateMemWit},\mathsf{VerifyMem})$ consists of the following five algorithms.

\begin{itemize}
	\item $\mathsf{Setup}(1^\lambda) \rightarrow pp$: Given a security parameter $\lambda$, this setup algorithm outputs a public parameter $pp$.
	\item $\mathsf{Commit}(pp, S) \rightarrow A^S$: Given the public parameter $pp$ and a set $S\subseteq D$, this committing algorithm outputs an accumulator digest $A^S$ to the set $S$.
	\item $\mathsf{Add}(A^S, ct) \rightarrow A^{S\cup \{ct\}}$: Given an accumulator digest $A^S$ to a set $S$ and an element $ct\in D\setminus S$, this adding algorithm outputs a new accumulator digest $A^{S\cup \{ct\}}$ to the set $S\cup \{ct\}$.
	\item $\mathsf{CreateMemWit}(S, A^S, ct) \rightarrow w^{ct, S}$: Given a set $S$, an accumulator digest $A^S$ to the set $S$, and an element $ct \in S$, this witness creation algorithm outputs the membership witness $w^{ct, S}$ of $ct \in S$.
	\item $\mathsf{VerifyMem}(A^S, ct, w^{ct, S}) \rightarrow b$: Given an accumulator digest $A^S$ to a set $S$, an element $ct$, and a membership witness $w^{ct, S}$, this membership verification algorithm outputs a bit $b\gets1$ to indicate that $w^{ct, S}$ is a valid witness for proving $ct\in S$; otherwise outputs $b\gets 0$.
\end{itemize}

\subsection{Zero-knowledge Proof} \label{Subsection_Zero-knowledgeProof}
The zero-knowledge proof (ZKP) scheme \cite{Paper_Nova} \cite{Paper_zkbridge} enables one to prove a statement without exposing other information.
The ZKP scheme $\mathcal{ZKP}=(\mathsf{Setup},\mathsf{Prove},\mathsf{Verify})$ consists of the following three algorithms.
\begin{itemize}
	\item $\mathsf{Setup}(1^\mu, R) \rightarrow crs$: Given a security parameter $\mu$ and a relationship $R$, this setup algorithm outputs a common reference string $crs$.
	\item $\mathsf{Prove}(crs, x, w) \rightarrow \pi$: Given a common reference string $crs$, a statement $x$, and a witness $w$, this proving algorithm outputs a proof $\pi$ for the relationship $R(x, w)$.
	\item $\mathsf{Verify}(crs, x, \pi) \rightarrow b$: Given a common reference string $crs$, the statement $x$, and the proof $\pi$, this verification algorithm outputs a bit 1/0 to indicate whether $R(x, w)$ holds or not.
\end{itemize}

ZKP satisfies three properties.
\emph{Completeness}: if the witness being proved is true, the verifier will be convinced of this fact with high probability.
\emph{Soundness}: if the witness being proved is false, no cheating prover can convince the verifier that it is true, except with a negligible probability.
\emph{Zero-knowledge}: the proof does not reveal any information about the witness being proved, except for the fact that it is true.

\section{Problem Statement} \label{Section_ProblemStatement}
\subsection{System Model} \label{Subsection_SystemModel}
\begin{figure}[tbp]
	\centering
	\includegraphics[width=1.00\linewidth]{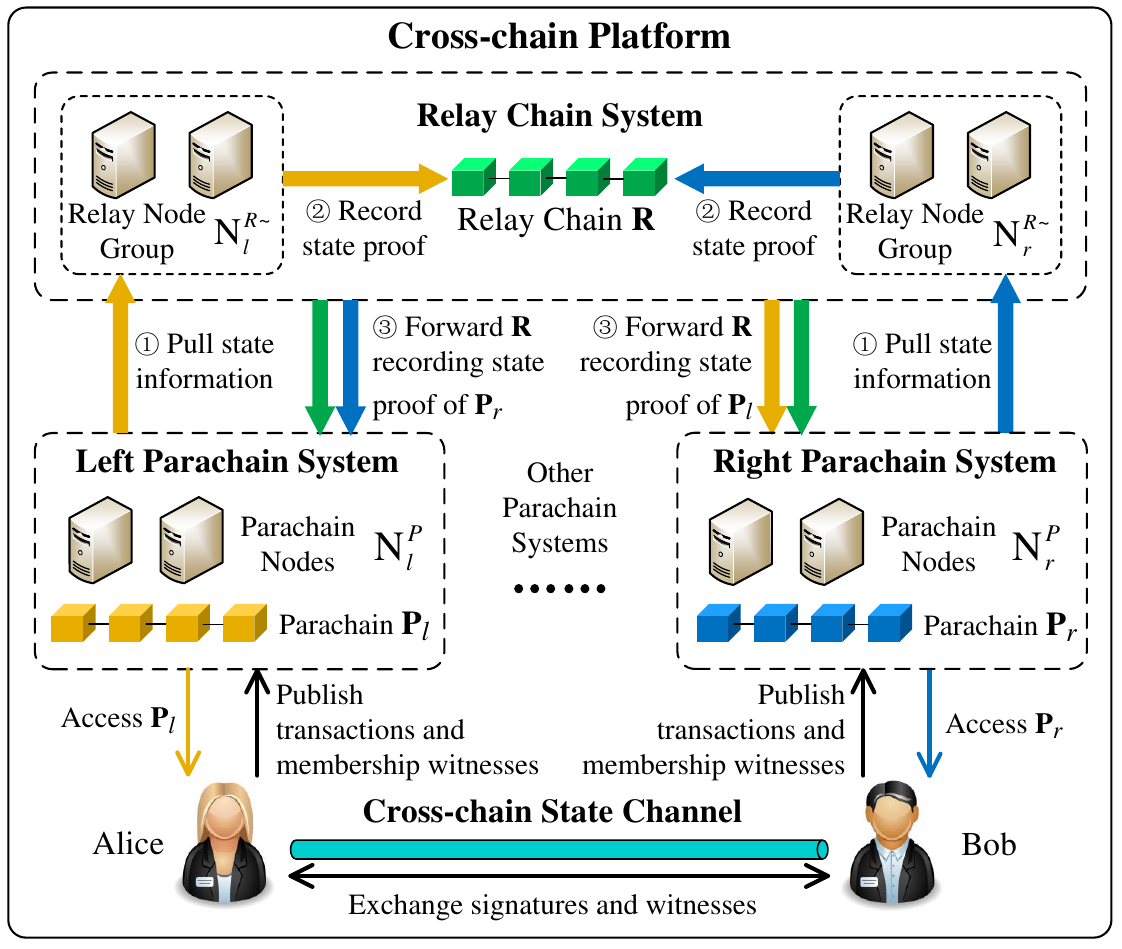}
	\caption{System model of Interpipe}
	\label{Figure_SystemModel}
\end{figure}

Interpipe includes one relay chain $\mathbf{R}$, and two parachains $\mathbf{P}_l$ and $\mathbf{P}_r$ called left parachain and right parachain. $\mathbf{P}_l$ and $\mathbf{P}_r$ have established cross-chain connections with $\mathbf{R}$ on a cross-chain platform, following relay-chain-parachain framework. The state information of $\mathbf{P}_l$ and $\mathbf{P}_r$ is synchronized into each other in the following way (see Fig. \ref{Figure_SystemModel}): \ding{172} $\mathrm{N}^{R \sim}_l$ and $\mathrm{N}^{R \sim}_r$ divided from $\mathrm{N}^R$ pull the latest state information of $\mathbf{P}_l$ and $\mathbf{P}_r$ from $\mathrm{N}^P_l$ and $\mathrm{N}^P_r$ using state pulling strategy; \ding{173} $\mathrm{N}^{R \sim}_l$ and $\mathrm{N}^{R \sim}_r$ generate the state proofs of $\mathbf{P}_l$ and $\mathbf{P}_r$, and record the state proofs into $\mathbf{R}$; \ding{174} as $\mathbf{R}$ is public on the cross-chain platform, $\mathrm{N}^P_r$ and $\mathrm{N}^P_l$ extract the $\mathbf{P}_l$ and $\mathbf{P}_r$ state proofs from $\mathbf{R}$, and record them into $\mathbf{P}_r$ and $\mathbf{P}_l$, achieving a synchronization.

There are two users Alice and Bob. On the one side, Alice belongs to $\mathbf{P}_l$ system. She can access the network of $\mathrm{N}^P_l$ to obtain the full blockchain data of $\mathbf{P}_l$, or publish new transactions on $\mathbf{P}_l$ via $\mathrm{N}^P_l$. Since the state proof of $\mathbf{P}_r$ has been recorded into $\mathbf{P}_l$, Alice can verify the state or transactions in both $\mathbf{P}_l$ and $\mathbf{P}_r$ by only accessing $\mathbf{P}_l$. In the same way, Bob belongs to $\mathbf{P}_r$ system, and he can verify the state or transactions in both $\mathbf{P}_l$ and $\mathbf{P}_r$ by only accessing $\mathbf{P}_r$. Additionally, a communication connection is established between Alice and Bob to transmit signatures and membership witnesses of transactions. This communication connection is off-chain, which does not require the use of any blockchain network. Based on the above configurations, a cross-chain state channel is established between Alice and Bob.

It is worth noting that among the $n$ parachains on the cross-chain platform, any two parachains can establish cross-chain state channels in the same way as parachain $\mathbf{P}_l$ and $\mathbf{P}_r$. For convenience, we specifically discuss a cross-chain state channel between $\mathbf{P}_l$ and $\mathbf{P}_r$ in the following content.

\subsection{Threat Model} \label{Subsection_ThreatModel}
\begin{itemize}
	\item \emph{Hard fork}: Hard fork can occur in a blockchain to result in a split from the original chain and the creation of a new separate chain.
	\item \emph{Denial-of-service attack}: With a sufficiently long period of time, an adversary can control every node in a relay node group to interrupt the information synchronization between parachains.
	\item \emph{Replay attack}: An adversary can replay cross-chain transactions and state proofs, attempting to have duplicate operations to cross-chain state channel.
	\item \emph{Counterfeiting}: An adversary can tamper the contents in cross-chain transactions and state proofs, attempting to synchronize false information between parachains.
	\item \emph{Eclipse attack}: An adversary can create a fake network environment around a user, attempting to prevent the user from learning the new states of blockchains.
	\item \emph{Conspiracy attack}: Alice and Bob can collude to publish different cross-chain transactions to $\mathbf{P}_l$ and $\mathbf{P}_r$, attempting to create parachain data out of thin air.
	\item \emph{Noncooperation}: One of Alice and Bob can refuse to cooperate with the other one, attempting to terminate the cross-chain transaction publication to $\mathbf{P}_l$ and $\mathbf{P}_r$.
\end{itemize}

We assume the cryptographic primitives, including hash function and digital signature, of relay chain and parachains are secure. Moreover, the zero-knowledge proof algorithms can be deployed in distributed environment without trusted setup \cite{Paper_Nova} \cite{Paper_zkbridge}.

As Interpipe establishes cross-chain state channels by the cooperation of three blockchains $\mathbf{R}$, $\mathbf{P}_l$, and $\mathbf{P}_r$, we assume the proportion of the corrupted consensus participators in each blockchain system is bounded by the threshold to ensure security. In the existing PoW \cite{Paper_SecurityOfPOWBlockchain} and PoS \cite{Paper_Ouroboros} protocol, it requires the proportion $\alpha < 1/3$.

Moreover, we assume that an adversary is mildly adaptive \cite{Paper_Ouroboros}. Specifically, for a group of honest nodes $\{node_1, \dots, node_x\}$, the adversary cannot instantly corrupt every node in $\{node_1, \dots, node_x\}$, and the corruption may only succeed after a sufficiently long period of time. Otherwise, the adversary possesses enough power to effortlessly control the majority of blockchain nodes, and overthrows the proportion $\alpha < 1/3$ or the security threshold in any node group.

Based on the above assumption, we can ensure persistence and liveness \cite{Paper_BitcoinBackbone} of blockchain transactions. Moreover, we ensure a stable block time, as the block generation process is controlled by mining difficulty \cite{Paper_DifficultyControl} in PoW consensus protocol or consensus slot \cite{Paper_Ouroboros} in PoS consensus protocol.

\begin{itemize}
	\item \emph{Transaction persistence} states that once a transaction goes more than $k$ blocks deep into the blockchain of one honest consensus participator, it will be included in every honest participator's blockchain with overwhelming probability.
	\item \emph{Transaction liveness} states that every transaction originating from an honest user will eventually end up at a depth of more than $k$ blocks in an honest consensus participator's blockchain, and an adversary cannot perform a selective denial-of-service attack against the honest user.
	\item \emph{Stable block time} states that the average time it takes for new blocks to be added to a blockchain remains consistent and predictable over an extended period.
\end{itemize}

\subsection{Design Goals} \label{Subsection_DesignGoals}
\begin{itemize}
	\item \emph{Consistency}: Parachain $\mathbf{P}_l$ and $\mathbf{P}_r$ can synchronize the real-time state proofs of each other. A cross-chain transaction $ct$ can be recorded into both $\mathbf{P}_l$ and $\mathbf{P}_r$, with their consistency being kept.
	\item \emph{Resistance}: It is hard for an adversary to interrupt the cross-chain synchronization between $\mathbf{P}_l$ and $\mathbf{P}_r$. Resistance relies only on the security guarantees in $\mathbf{R}$, $\mathbf{P}_l$, and $\mathbf{P}_r$ systems.
	\item \emph{Liveness}: Any two users respectively located in $\mathbf{P}_l$ and $\mathbf{P}_r$ systems can have opening, updating, closing, and disputing to a cross-chain state channel.
	\item \emph{Efficiency}: Any user in $\mathbf{R}$, $\mathbf{P}_l$, and $\mathbf{P}_r$ systems can have cross-chain verification to the state and transactions in $\mathbf{R}$, $\mathbf{P}_l$ and $\mathbf{P}_r$ with low storage and computation overhead.
\end{itemize}

\section{Batch Transaction Proof} \label{Section_BatchTransactionProof}
Batch transaction proof enables verifiers in $\mathbf{P}_l$ or $\mathbf{P}_r$ system to have verification to any cross-chain transaction in $\mathbf{P}_r$ or $\mathbf{P}_l$ with low storage and computation overhead. This scheme has three processes, which are initialization, recursive proving, and verification.

\textbf{Initialization}: The initialization is executed when relay chain $\mathbf{R}$ and parachains $\mathbf{P}_i$ $(i \in [n])$ first establish cross-chain connections on cross-chain platform. Specifically, the blockchain protocols of relay chain and parachains including data structure, encryption algorithm, and consensus mechanism are published to every entity on the cross-chain platform. Second, the identity information including public keys and blockchain addresses of blockchain nodes $\mathrm{N}^R$, $\mathrm{N}^P_i$ $(i \in [n])$ are also published on the cross-chain platform, which will be used as the public statement $x$ to verify new blocks and state proofs following the scheme in \cite{Paper_Nova}. Moreover, the blockchain nodes generate public parameter $pp \leftarrow \mathcal{ACC}.\mathsf{Setup}(1^\lambda)$ and $crs \leftarrow \mathcal{ZKP}.\mathsf{Setup}(1^\mu, R)$ for accumulator and zero knowledge proof in distributed environment.

Then, the group $\mathrm{N}^{R \sim}_l$ is divided from relay node set $\mathrm{N}^R$ to establish cross-chain connections with $\mathrm{N}^P_l$, and pull the data of $\mathbf{P}_l$ from $\mathrm{N}^P_l$. $\mathrm{N}^{R \sim}_l$ generate the initial accumulator $A^l_0$, which can be the generator without adding elements. Moreover, $\mathrm{N}^{R \sim}_l$ generate the zero knowledge proof $\pi^l_0$ of $\mathbf{P}_l$, which can be achieved by the existing scheme \cite{Paper_zkbridge}. The tuple $(A^l_0, \pi^l_0)$ is called the initial state proof of $\mathbf{P}_l$.

Noted, in the following content, we only illustrate the batch transaction proof of $\mathbf{P}_l$ for convenience, while the proof of $\mathbf{P}_r$ is generated by the same method of $\mathbf{P}_l$.

\begin{figure}[tbp]
	\centering
	\includegraphics[width=1.00\linewidth]{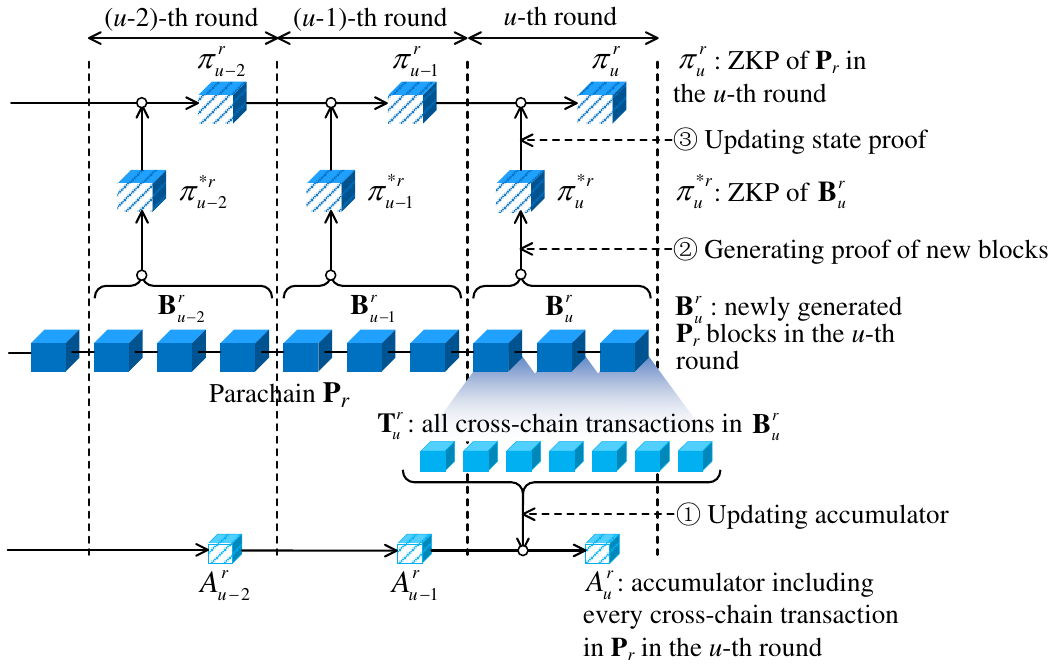}
	\caption{Recursive proof of parachain}
	\label{Figure_RecursiveProofOfParachain}
\end{figure}

\textbf{Recursive proving}: Recursive proving is executed in rounds, which are a continuous series of time intervals with fixed length. The proving process in the $u$-th round includes three steps, which are updating accumulator, generating proof of new block, and updating state proof. The proof generation process is shown in Fig. \ref{Figure_RecursiveProofOfParachain}.

\ding{172} Updating accumulator. In the $u$-th round, there are multiple new $\mathbf{P}_l$ blocks generated, noted as $\mathbf{B}^l_u$. The relay node group $\mathrm{N}^{R \sim}_l$ pull $\mathbf{B}^l_u$ from $\mathrm{N}^P_l$, extract all cross-chain transactions $\mathbf{T}^l_u$ from $\mathbf{B}^l_u$, and add $\mathbf{T}^l_u$ into accumulator $A^l_{u-1}$ to generate $A^l_u \leftarrow \mathcal{ACC}.\mathsf{Add}(A^l_{u-1}, \mathbf{T}^l_u)$, where $A^l_{u-1}$ is the accumulator in the ($u$-1)-th round. Noted, the ordinary intra-chain transactions in $\mathbf{B}^l_u$ which do not need to be cross-chain synchronized will not be extracted or added into the accumulator.

\ding{173} Generating proof of new blocks. $\mathrm{N}^{R \sim}_l$ generate the zero-knowledge proof $\pi^{*r}_u \leftarrow \mathcal{ZKP}.\mathsf{Prove}(crs, x, A^l_{u-1}, A^l_u, \mathbf{B}^l_u)$. The witness to be proved includes the following.

\begin{itemize}
	\item There is a set of new parachain blocks $\mathbf{B}^l_u$. Every block in $\mathbf{B}^l_u$ has the correct format with valid proof of work (in PoW protocol) or proof of stake (in PoS protocol). Every block in $\mathbf{B}^l_u$ has a valid hash pointer pointing to the last block.
	\item Every cross-chain transaction in $\mathbf{B}^l_u$ has been correctly included in the Merkle tree of the block in $\mathbf{B}^l_u$.
	\item Every cross-chain transaction in $\mathbf{B}^l_u$ has been added in $A^l_{u-1}$ to output $A^l_u$.
\end{itemize}

The above witness will be transformed into arithmetic circuits to be substituted into the proof process of zk-SNARK to output the proof $\pi^{*r}_u$.

\ding{174} Updating state proof. We have a recursive proof by using the recursive SNARK in Nova \cite{Paper_Nova} to further generate $\pi^l_u \leftarrow \mathcal{ZKP}.\mathsf{Prove}(crs, x, \pi^l_{u-1}, \pi^{*l}_u)$, in which $\pi^l_{u-1}$ is the proof generated in the ($u$-1)-th round. By this step, the zero knowledge proof of $\mathbf{P}_l$ is updated from $\pi^l_{u-1}$ to $\pi^l_u$. Subsequently, we call $(\pi^l_u, A^l_u)$ as the state proof of $\mathbf{P}_l$ in the $u$-th round. More importantly, to obtain $\pi^l_u$, the prover only needs to calculate the arithmetic circuits in the new blocks $\mathbf{B}^l_u$ and the proof $\pi^l_{u-1}$ in the last round. The old blocks $\mathbf{B}^l_x$ $(0 < x < u)$ generated in the previous rounds do not need to be recalculated, as the witness of $\mathbf{B}^l_x$ have been proved by $\pi^l_{u-1}$. Because the number of new blocks in a round is relatively small, the calculation to the arithmetic circuits can be completed in a short time.

\textbf{Verification}: On the side of $\mathbf{P}_r$, without accessing the original data of $\mathbf{P}_l$, verifiers will confirm the correctness of $(\pi^l_u, A^l_u)$ by $\mathcal{ZKP}.\mathsf{Verify}(crs, x, A^l_{u-1}, A^l_u, \pi^l_u)$. Specifically, there is a valid parachain $\mathbf{P}_l$, and every cross-chain transaction in $\mathbf{P}_l$ has been added into $A^l_u$. The identities of the verifiers are $\mathbf{P}_r$ node set $\mathrm{N}^P_r$ and any user in $\mathbf{P}_r$ system including Alice.

Based on $A^l_u$, verifiers in $\mathbf{P}_r$ system can further verify that a certain cross-chain transaction $ct$ has been recorded in $\mathbf{P}_l$. This requires a prover connecting with $\mathbf{P}_l$ system to create the membership witness $w^{ct, l}_u$ of $ct$ in $A^l_u$, where $w^{ct, l}_u \leftarrow \mathcal{ACC}.\mathsf{CreateMemWit}(\mathbf{P}_l, A^l_u, ct)$, and send $w^{ct, l}_u$ to the verifiers. Based on $w^{ct, l}_u$, the verifiers have verification by $\mathcal{ACC}.\mathsf{VerifyMem}(A^l_u, ct, w^{ct, l}_u)$ to confirm that $ct$ has been added into $A^l_u$, and consequently, $ct$ has been recorded in $\mathbf{P}_l$.

By the above method, verifiers first need to have one verification to $(\pi^l_u, A^l_u)$ to confirm the correctness of $A^l_u$. Then, based on $A^l_u$, the verifiers can verify any cross-chain transactions in $\mathbf{P}_l$. Each verification of a cross-chain transaction only requires one calculation of $\mathcal{ACC}.\mathsf{VerifyMem}(A^l_u, ct, w^{ct, l}_u)$, which can satisfy the demands of scalable cross-chain transaction verification.

\section{Interpipe} \label{Section_Interpipe}
In this section, we first have an illustration of cross-chain synchronization (see Section \ref{Subsection_Cross-chainSynchronization}), which enables two parachains $\mathbf{P}_l$ and $\mathbf{P}_r$ to have consistent operations. Based on cross-chain synchronization, we provide designs for the cross-chain state channel operations of opening, updating, closing, and disputing (see Section \ref{Subsection_Cross-chainStateChannel}).

\subsection{Cross-chain Synchronization} \label{Subsection_Cross-chainSynchronization}
Cross-chain synchronization has two parts, which are state synchronization and transaction synchronization. First, state synchronization is the underlying design. It is recursively executed in rounds, enabling two parachains to keep real-time records of the state proofs of each other. Based on the real-time state proofs, we achieve transaction synchronization. It means that users can record one cross-chain transaction $ct$ into the two parachains $\mathbf{P}_l$ and $\mathbf{P}_r$, and keep their consistency.

\subsubsection{State synchronization}

\begin{figure*}[t]
	\centering
	\includegraphics[width=0.65\linewidth]{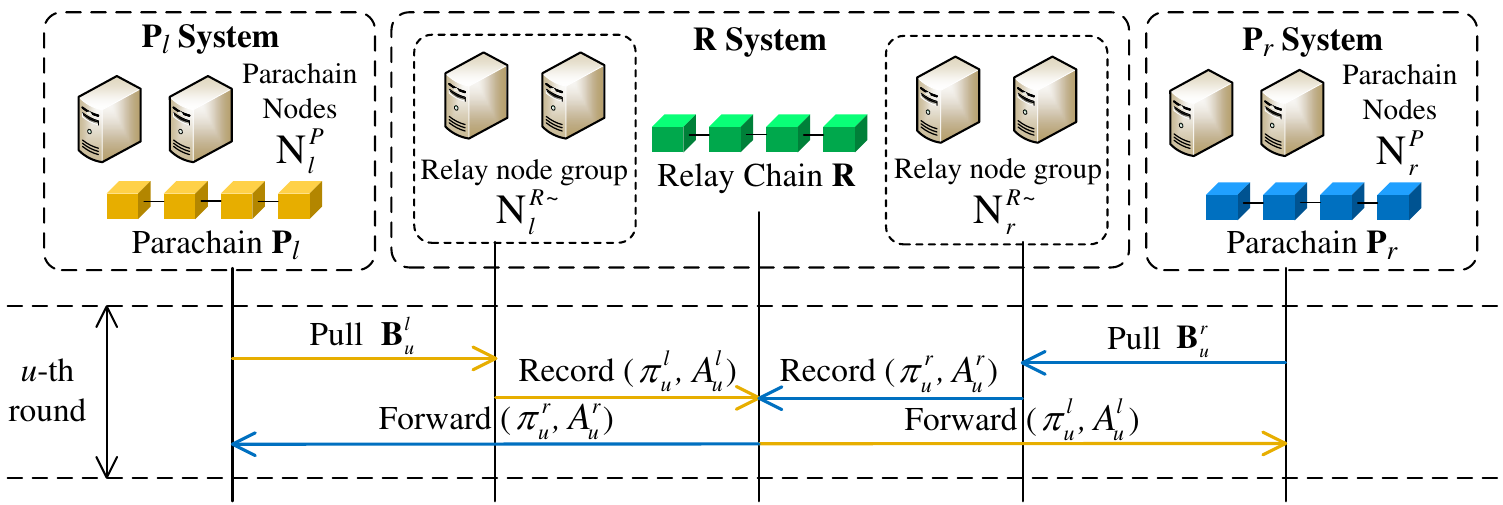}
	\caption{One round of state synchronization}
	\label{Figure_OneRoundOfStateSynchronization}
\end{figure*}

First, state synchronization follows the initialization of batch transaction proof to generate the initial state proof $(A^l_0, \pi^l_0)$ of $\mathbf{P}_l$ (see line 2-11 in Protocol \ref{Protocol_StateSynchronization}). Then, $\mathrm{N}^{R \sim}_l$ record $(A^l_0, \pi^l_0)$ into $\mathbf{R}$ to start state synchronization. Because $\mathbf{P}_l$ continues to grow over time, the $\mathbf{P}_l$ state keeps updating. To keep the real-time $\mathbf{P}_l$ state proof synchronized into $\mathbf{P}_r$, the state synchronization repeats at regular time intervals called rounds (see line 12-17 in Protocol \ref{Protocol_StateSynchronization}). In the $u$-th round, the state information of $\mathbf{P}_l$ and $\mathbf{P}_r$ will be transferred and recorded into each other based on cross-chain state transfer \cite{Paper_Polkadot} \cite{Paper_Cosmos} with two processes of state reception and state forwarding. In the following content, we further design the two processes, and illustrate the information transfer from $\mathbf{P}_l$ to $\mathbf{P}_r$, where the information transfer from $\mathbf{P}_r$ to $\mathbf{P}_l$ follows the same way.

In state reception process (see line 18-25 in Protocol \ref{Protocol_StateSynchronization}), $\mathrm{N}^P_l$ transfer the state information of $\mathbf{P}_l$ to $\mathrm{N}^{R \sim}_l$. The state information is in the form of newly generated $\mathbf{P}_l$ blocks in the $u$-th round, noted as $\mathbf{B}^l_u$. Then, $\mathrm{N}^{R \sim}_l$ extract all cross-chain transactions $\mathbf{T}^l_u$ which need to be cross-chain synchronized from $\mathbf{B}^l_u$, and add $\mathbf{T}^l_u$ into $A^l_{u-1}$ to generate $A^l_u$, where $A^l_{u-1}$ is the accumualtor in the ($u$-1)-th round. Next, $\mathrm{N}^{R \sim}_l$ generate the zero-knowledge proof $\pi^l_u$ of the current $\mathbf{P}_l$ by recursive proving (see Section \ref{Section_BatchTransactionProof}), where $A^l_u$ is binded with $\pi^l_u$ to have $(\pi^l_u, A^l_u)$ as the state proof of $\mathbf{P}_l$. Then, $\mathrm{N}^{R \sim}_l$ will record $(\pi^l_u, A^l_u)$ into relay chain $\mathbf{R}$ to be published on cross-chain platform.

Moreover, in the $u$-th round, $\mathrm{N}^{R \sim}_l$ use state pulling strategy \cite{Paper_XPull} to pull $\mathbf{B}^l_u$ from $\mathrm{N}^P_l$. Therefore, every time there are new blocks generated in $\mathbf{P}_l$, the new blocks will be obtained by $\mathrm{N}^{R \sim}_l$ within a certain number of rounds. Subsequently, the time between the moment in which new blocks are generated and the moment in which new state proof is recorded into $\mathbf{P}_r$ will not exceed a certain limit, which ensures the timeliness of state synchronization.

\begin{algorithm}[t]
	\footnotesize
	\setstretch{1.1}
	\caption{State Synchronization}
	\label{Protocol_StateSynchronization}
	
	\Fn{StateSynchronization}{
		// Initialization\;
		\begin{tabular}{@{}p{2.34cm}@{}l@{}} \raggedright             $\mathrm{N}^R$, $\mathrm{N}^P_i$ $(i \in [n])$ &
			: publish blockchain protocols\\
			&
			~~and identity information\\																												\end{tabular}\;
		\begin{tabular}{@{}p{2.34cm}@{}l@{}} \raggedright             $\mathrm{N}^R$, $\mathrm{N}^P_i$ $(i \in [n])$ &
			: $\mathcal{ZKP}.\mathsf{Setup}$, $\mathcal{ACC}.\mathsf{Setup}$, $u \leftarrow 1$\\														\end{tabular}\;
		\For{$i \gets l, r$}{
			\begin{tabular}{@{}p{1.8cm}@{}l@{}} \raggedright	   $\mathrm{N}^R \rightarrow \mathrm{N}^{R \sim}_i$ &
				: $\mathrm{N}^R$ divide groups\\																										\end{tabular}\;
			\begin{tabular}{@{}p{1.8cm}@{}l@{}} \raggedright $\mathrm{N}^{R \sim}_i \Leftrightarrow \mathrm{N}^P_i$ &
				: $\mathrm{N}^{R \sim}_i$ establish connections with $\mathrm{N}^P_i$\\																	\end{tabular}\;
			\begin{tabular}{@{}p{1.8cm}@{}l@{}} \raggedright      $\mathrm{N}^{R \sim}_i \Leftarrow \mathrm{N}^P_i$ &
				: $\mathrm{N}^{R \sim}_i$ pull $\mathbf{P}_i$ from $\mathrm{N}^P_i$ \\																	\end{tabular}\;
			\begin{tabular}{@{}p{0.8cm}@{}l@{}} \raggedright                                $\mathrm{N}^{R \sim}_i$ &
				: $\mathrm{N}^{R \sim}_i$ set the initial accumulator $A^i_0$\\																			\end{tabular}\;
			\begin{tabular}{@{}p{0.8cm}@{}l@{}} \raggedright                                $\mathrm{N}^{R \sim}_i$ &
				: $\mathcal{ZKP}.\mathsf{Prove}(crs, x, \mathbf{P}_i, A^i_0) \rightarrow \pi^i_0$\\														\end{tabular}\;
			\begin{tabular}{@{}p{0.8cm}@{}l@{}} \raggedright                                $\mathrm{N}^{R \sim}_i$ &
				: $\mathrm{N}^{R \sim}_i$ record $(\pi^i_0, A^i_0)$ into $\mathbf{R}$\\																	\end{tabular}\;
		}
		
		\vspace{0.5em}
		// Synchronization in rounds\;
		\For{true}{
			\For{$i \gets l, r$}{
				\begin{tabular}{@{}p{0.8cm}@{}l@{}} \raggedright $\mathrm{N}^{R \sim}_i$ &
					: $\mathrm{N}^{R \sim}_i$ have \emph{StateReception}\\																				\end{tabular}\;
				\begin{tabular}{@{}p{0.8cm}@{}l@{}} \raggedright        $\mathrm{N}^P_j$ &
					: $\mathrm{N}^P_j$ have \emph{StateForwarding}\\																					\end{tabular}\;
				\begin{tabular}{@{}p{0.8cm}@{}l@{}} \raggedright          $\mathrm{N}^R$ &
					: Round number $u$ increases by one\\																								\end{tabular}\;
			}
		}
	}
	
	\vspace{0.5em}
	\Fn{StateReception}{
		\For{$i \gets l, r$}{
			\begin{tabular}{@{}p{1.8cm}@{}l@{}} \raggedright $\mathrm{N}^{R \sim}_i \Leftarrow \mathrm{N}^P_i$ &
				: $\mathrm{N}^{R \sim}_i$ pull the new parachain blocks $\mathbf{B}^i_u$ \\
				&
				~~generated in the $u$-th round from $\mathrm{N}^P_i$\\																					\end{tabular}\;
			\begin{tabular}{@{}p{0.8cm}@{}l@{}} \raggedright                           $\mathrm{N}^{R \sim}_i$ &
				: $\mathrm{N}^{R \sim}_i$ extract all cross-chain trans $\mathbf{T}^i_u$ from $\mathbf{B}^i_u$\\										\end{tabular}\;
			\begin{tabular}{@{}p{0.8cm}@{}l@{}} \raggedright                           $\mathrm{N}^{R \sim}_i$ &
				: $\mathcal{ACC}.\mathsf{Add}(A^i_{u-1}, \mathbf{T}^i_u) \rightarrow A^i_u$\\															\end{tabular}\;
			\begin{tabular}{@{}p{0.8cm}@{}l@{}} \raggedright                           $\mathrm{N}^{R \sim}_i$ &
				: $\mathcal{ZKP}.\mathsf{Prove}(crs, x, A^i_{u-1}, A^i_u, \mathbf{B}^i_u) \rightarrow \pi^{*i}_u$\\										\end{tabular}\;
			\begin{tabular}{@{}p{0.8cm}@{}l@{}} \raggedright                           $\mathrm{N}^{R \sim}_i$ &
				: $\mathcal{ZKP}.\mathsf{Prove}(crs, x, \pi^i_{u-1}, \pi^{*i}_u) \rightarrow \pi^i_u$\\													\end{tabular}\;
			\begin{tabular}{@{}p{0.8cm}@{}l@{}} \raggedright                           $\mathrm{N}^{R \sim}_i$ &
				: $\mathrm{N}^{R \sim}_i$ record $(\pi^i_u, A^i_u)$ into $\mathbf{R}$\\																	\end{tabular}\;
		}
	}
	
	\vspace{0.5em}
	\Fn{StateForwarding}{
		\For{$i \gets l, r$ and $j \gets r, l$}{
			\begin{tabular}{@{}p{0.8cm}@{}l@{}} \raggedright $\mathrm{N}^P_j$ & : $\mathrm{N}^P_j$ extract $(\pi^i_u, A^i_u)$ from $\mathbf{R}$\\		\end{tabular}\;
			\begin{tabular}{@{}p{0.8cm}@{}l@{}} \raggedright $\mathrm{N}^P_j$ & : $\mathcal{ZKP}.\mathsf{Verify}(crs, x, \pi^i_u, A^i_u)$\\				\end{tabular}\;
			\begin{tabular}{@{}p{0.8cm}@{}l@{}} \raggedright $\mathrm{N}^P_j$ & : $\mathrm{N}^P_j$ record $(\pi^i_u, A^i_u)$ into $\mathbf{P}_j$\\		\end{tabular}\;
		}
	}
\end{algorithm}

In state forwarding process (see line 26-30 in Protocol \ref{Protocol_StateSynchronization}), $\mathbf{R}$ is published to $\mathbf{P}_r$ system to reach consensus among $\mathrm{N}^P_r$. Then, $\mathrm{N}^P_r$ extract $(\pi^l_u, A^l_u)$ from $\mathbf{R}$, verify the correctness of $(\pi^l_u, A^l_u)$, and record $(\pi^l_u, A^l_u)$ into $\mathbf{P}_r$. We say that the $\mathbf{P}_l$ state proof $(\pi^l_u, A^l_u)$ is synchronized into $\mathbf{P}_r$. Furthermore, based on $(\pi^l_u, A^l_u)$, $\mathbf{P}_r$ system can verify that a certain cross-chain transaction $ct$ has been recorded into $\mathbf{P}_l$, with the help of a prover in $\mathbf{P}_l$ system to provide the membership witness $w^{ct, l}_u$ of $ct$.

Using the same method of state reception and state forwarding, $(\pi^r_u, A^r_u)$ is generated by $\mathrm{N}^{R \sim}_r$ to be recorded into $\mathbf{R}$, and then $(\pi^r_u, A^r_u)$ is extracted by $\mathrm{N}^P_l$ to be recorded into $\mathbf{P}_l$. Therefore, in the same round, $\mathbf{P}_l$ and $\mathbf{P}_r$ have their state proofs synchronized to each other. Fig. \ref{Figure_OneRoundOfStateSynchronization} shows one round of state synchronization.

\subsubsection{Transaction synchronization}
Based on state synchronization, we further achieve transaction synchronization to record one cross-chain transaction $ct$ into two parachains $\mathbf{P}_l$ and $\mathbf{P}_r$, and keep their consistency. Classified by the initiator, transaction synchronization can be divided into two categories, which are unilaterally initiated transaction synchronization (UITS) and jointly initiated transaction synchronization (JITS).

\begin{figure*}[t]
	\centering
	\includegraphics[width=0.9\linewidth]{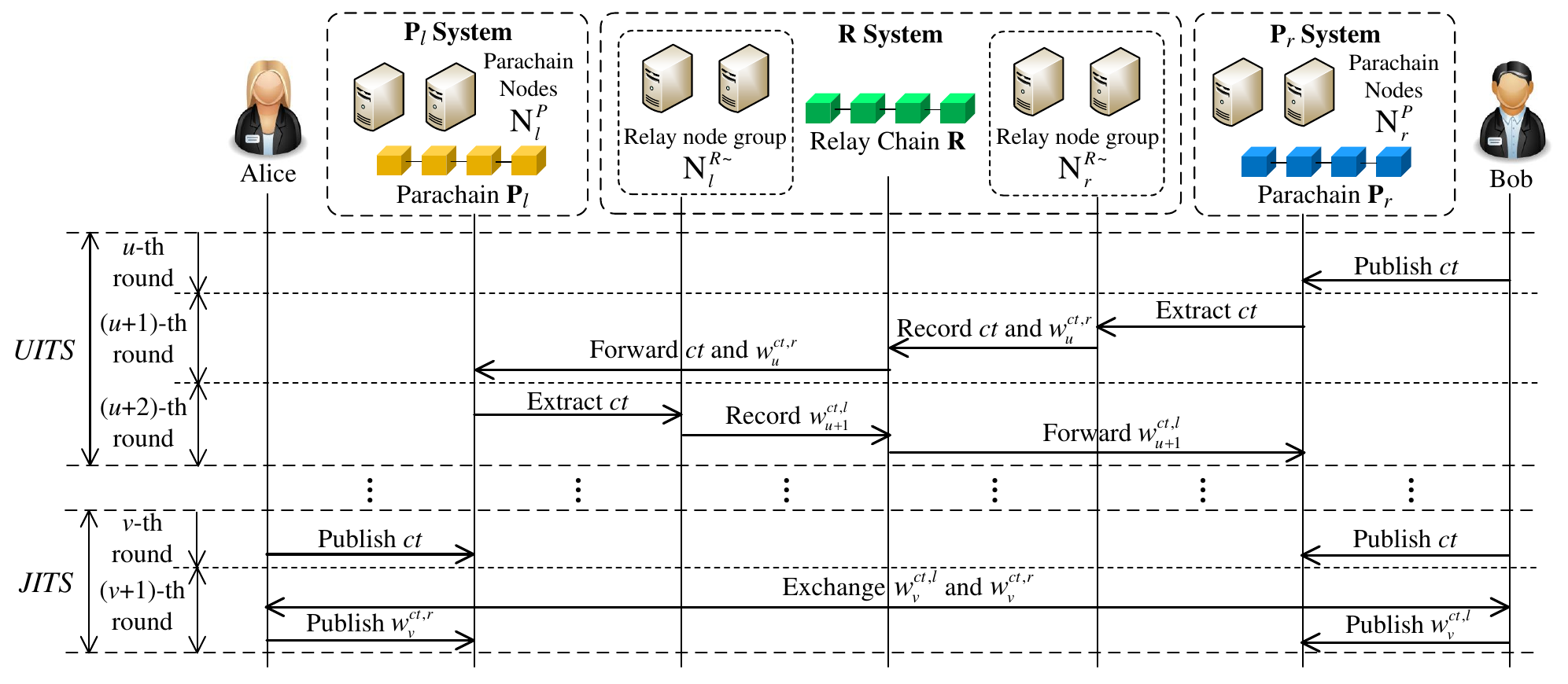}
	\caption{Transaction synchronization}
	\label{Figure_TransactionSynchronization}
\end{figure*}

\begin{algorithm} [t]
	\footnotesize
	\setstretch{1.1}
	\caption{Transaction Synchronization}
	\label{Protocol_TransactionSynchronization}
	
	\Fn{UITS}{
		// $i \gets l$ or $r$~~~// $j \gets r$ or $l$~~~// $U^i \leftarrow$ Alice or Bob\;
		// $u$-th round\;
		\begin{tabular}{@{}p{1.4cm}@{}l@{}} \raggedright  			  $U^i \Rightarrow \mathrm{N}^P_i$ &
			: $U^i$ publishes $ct$ to $\mathrm{N}^P_i$\\																								\end{tabular}\;
		\begin{tabular}{@{}p{1.4cm}@{}l@{}} \raggedright                              $\mathrm{N}^P_i$ &
			: $\mathrm{N}^P_i$ record $ct$	into $\mathbf{P}_i$\\																						\end{tabular}\;
		
		\vspace{0.5em}
		// ($u$+1)-th round\;
		\begin{tabular}{@{}p{1.4cm}@{}l@{}} \raggedright                       $\mathrm{N}^{R \sim}_i$ &
			: $\mathrm{N}^{R \sim}_i$ extract $ct$ from $\mathbf{P}_i$\\																				\end{tabular}\;
		\begin{tabular}{@{}p{1.4cm}@{}l@{}} \raggedright                       $\mathrm{N}^{R \sim}_i$ &
			: $\mathcal{ACC}.\mathsf{CreateMemWit}(\mathbf{P}_i, A^i_u, ct) \rightarrow w^{ct, i}_u$\\													\end{tabular}\;
		\begin{tabular}{@{}p{1.4cm}@{}l@{}} \raggedright                       $\mathrm{N}^{R \sim}_i$ &
			: $\mathrm{N}^{R \sim}_i$ record $(ct, w^{ct, i}_u)$ into $\mathbf{R}$\\																	\end{tabular}\;
		\begin{tabular}{@{}p{1.4cm}@{}l@{}} \raggedright                              $\mathrm{N}^P_j$ &
			: $\mathrm{N}^P_j$ extract $(ct, w^{ct, i}_u)$ from $\mathbf{R}$\\																			\end{tabular}\;
		\begin{tabular}{@{}p{1.4cm}@{}l@{}} \raggedright                              $\mathrm{N}^P_j$ &
			: $\mathcal{ACC}.\mathsf{VerifyMem}(A^i_u, ct, w^{ct, i}_u)$\\																				\end{tabular}\;
		\begin{tabular}{@{}p{1.4cm}@{}l@{}} \raggedright                              $\mathrm{N}^P_j$ &
			: $\mathrm{N}^P_j$ record $(ct, w^{ct, i}_u)$ into $\mathbf{P}_j$\\																			\end{tabular}\;
		
		\vspace{0.5em}
		// ($u$+2)-th round\;
		\begin{tabular}{@{}p{1.4cm}@{}l@{}} \raggedright                       $\mathrm{N}^{R \sim}_j$ &
			: $\mathrm{N}^{R \sim}_j$ extract $ct$ from $\mathbf{P}_j$\\																				\end{tabular}\;
		\begin{tabular}{@{}p{1.4cm}@{}l@{}} \raggedright                       $\mathrm{N}^{R \sim}_j$ &
			: $\mathcal{ACC}.\mathsf{CreateMemWit}(\mathbf{P}_j, A^j_{u+1}, ct) \rightarrow w^{ct, j}_{u+1}$\\											\end{tabular}\;
		\begin{tabular}{@{}p{1.4cm}@{}l@{}} \raggedright                       $\mathrm{N}^{R \sim}_j$ &
			: $\mathrm{N}^{R \sim}_j$ record $w^{ct, j}_{u+1}$ into $\mathbf{R}$\\																		\end{tabular}\;
		\begin{tabular}{@{}p{1.4cm}@{}l@{}} \raggedright                              $\mathrm{N}^P_i$ &
			: $\mathrm{N}^P_i$ extract $w^{ct, j}_{u+1}$ from $\mathbf{R}$\\																			\end{tabular}\;
		\begin{tabular}{@{}p{1.4cm}@{}l@{}} \raggedright                              $\mathrm{N}^P_i$ &
			: $\mathcal{ACC}.\mathsf{VerifyMem}(A^j_{u+1}, ct, w^{ct, j}_{u+1})$\\																		\end{tabular}\;
		\begin{tabular}{@{}p{1.4cm}@{}l@{}} \raggedright                              $\mathrm{N}^P_i$ &
			: $\mathrm{N}^P_i$ record $w^{ct, j}_{u+1}$ into $\mathbf{P}_i$\\																			\end{tabular}\;
	}
	
	\vspace{0.5em}
	\Fn{JITS}{
		// $U^l \leftarrow$ Alice ~~~// $U^r \leftarrow$ Bob\;
		// $v$-th round\;
		\begin{tabular}{@{}p{3cm}@{}l@{}} \raggedright $U^l$ / $U^r$ $\Rightarrow$ $\mathrm{N}^P_l$ / $\mathrm{N}^P_r$&
			: $U^l$ / $U^r$ publish $ct$ to $\mathrm{N}^P_l$ / $\mathrm{N}^P_r$\\																		\end{tabular}\;
		\begin{tabular}{@{}p{1.4cm}@{}l@{}} \raggedright        			      $\mathrm{N}^P_l$ / $\mathrm{N}^P_r$ &
			: $\mathrm{N}^P_l$ / $\mathrm{N}^P_r$ record $ct$ into $\mathbf{P}_l$ / $\mathbf{P}_r$\\													\end{tabular}\;
		
		\vspace{0.5em}
		// ($v$+1)-th round\;
		\begin{tabular}{@{}p{1.4cm}@{}l@{}} \raggedright            								    $U^l$ / $U^r$ &
			: $\mathcal{ACC}.\mathsf{CreateMemWit}(\mathbf{P}_l, A^l_v, ct) \rightarrow w^{ct, l}_v$ / \\
			&
			~~$\mathcal{ACC}.\mathsf{CreateMemWit}(\mathbf{P}_r, A^r_v, ct) \rightarrow w^{ct, r}_v$\\													\end{tabular}\;
		\begin{tabular}{@{}p{1.4cm}@{}l@{}} \raggedright 									$U^l \Leftrightarrow U^r$ &
			: $U^l$ and $U^r$ exchange $w^{ct, l}_v$ and $w^{ct, r}_v$\\																				\end{tabular}\;
		\begin{tabular}{@{}p{1.4cm}@{}l@{}} \raggedright			 								    $U^l$ / $U^r$ &
			: $U^l$ / $U^r$ publish $w^{ct, r}_v$ / $w^{ct, l}_v$ to $\mathrm{N}^P_l$ / $\mathrm{N}^P_r$\\												\end{tabular}\;
		\begin{tabular}{@{}p{1.4cm}@{}l@{}} \raggedright 						  $\mathrm{N}^P_l$ / $\mathrm{N}^P_r$ &
			: $\mathcal{ACC}.\mathsf{VerifyMem}(A^r_v, ct, w^{ct, r}_v)$ / \\
			&
			~~$\mathcal{ACC}.\mathsf{VerifyMem}(A^l_v, ct, w^{ct, l}_v)$\\																				\end{tabular}\;
		\begin{tabular}{@{}p{1.4cm}@{}l@{}} \raggedright          				  $\mathrm{N}^P_l$ / $\mathrm{N}^P_r$ &
			: $\mathrm{N}^P_l$ / $\mathrm{N}^P_r$ record $w^{ct, r}_v$ / $w^{ct, l}_v$ into $\mathbf{P}_l$ / $\mathbf{P}_r$\\							\end{tabular}\;
	}
\end{algorithm}

\textbf{UITS}: UITS is initiated by any single party belonging to $\mathbf{P}_l$ system or $\mathbf{P}_r$ system. For example, if Bob in $\mathbf{P}_r$ system wants to unilaterally record $ct$ into $\mathbf{P}_l$ and $\mathbf{P}_r$, he needs to go through the following steps.

\begin{enumerate}
	\item In the $u$-th round of state synchronization, Bob attaches a UITS label to $ct$, and publishes $ct$ to $\mathrm{N}^P_r$. Then, $\mathrm{N}^P_r$ record $ct$ into $\mathbf{P}_r$. (see line 3-5 in Protocol \ref{Protocol_TransactionSynchronization})
	
	\item In the ($u$+1)-th round of state synchronization, $\mathrm{N}^{R \sim}_r$ additionally generate the membership witness $w^{ct, r}_u$ of $ct$ in $\mathbf{P}_r$. $ct$ and $w^{ct, r}_u$ will be recorded into $\mathbf{R}$. Then, $ct$ and $w^{ct, r}_u$ are extracted by $\mathrm{N}^P_l$ to be recorded into $\mathbf{P}_l$. Subsequently, by verifying $ct$, $w^{ct, r}_u$, and $(\pi^r_u, A^r_u)$, $\mathbf{P}_l$ system will confirm that $ct$ has been recorded in $\mathbf{P}_r$. (see line 6-12 in Protocol \ref{Protocol_TransactionSynchronization})
	
	\item In the ($u$+2)-th round of state synchronization, $\mathrm{N}^{R \sim}_l$ additionally generate the membership witness $w^{ct, l}_{u+1}$ of $ct$ in $\mathbf{P}_l$. $w^{ct, l}_{u+1}$ will be recorded into $\mathbf{R}$, Then, $w^{ct, l}_{u+1}$ is extracted by $\mathrm{N}^P_r$ to be recorded into $\mathbf{P}_r$. Subsequently, by verifying $w^{ct, l}_{u+1}$ and $(\pi^l_{u+1}, A^l_{u+1})$, $\mathbf{P}_r$ system will confirm that $ct$ has been recorded in $\mathbf{P}_l$. (see line 13-19 in Protocol \ref{Protocol_TransactionSynchronization})
\end{enumerate}

It is worth noting that Alice does not participate in UITS with Bob, but she can monitor that $ct$ is recorded in $\mathbf{P}_l$, as $\mathbf{P}_l$ is public to the members in $\mathbf{P}_l$ system. The monitoring will enable Alice to learn the malicious behavior of Bob when he publishes an outdated transaction by UITS. This feature will be applied in the disputing operation to cross-chain state channel (see Section \ref{Subsection_Cross-chainStateChannel}).

\textbf{JITS}: If Alice and Bob intend to jointly record $ct$ into $\mathbf{P}_l$ and $\mathbf{P}_r$ by using JITS, they need to go through the following steps. To facilitate the description, the roles on both sides of the slash ``/" will perform the operation simultaneously in the following content.

\begin{enumerate}
	\item In the $v$-th round of state synchronization, Alice / Bob attaches a JITS label to $ct$, and publishes $ct$ to $\mathrm{N}^P_l$ / $\mathrm{N}^P_r$. Then, $\mathrm{N}^P_l$ / $\mathrm{N}^P_r$ records $ct$ into $\mathbf{P}_l$ / $\mathbf{P}_r$. (see line 22-24 in Protocol \ref{Protocol_TransactionSynchronization})
	
	\item In the ($v$+1)-th round of state synchronization, Alice / Bob generates the membership witness $w^{ct, l}_v$ / $w^{ct, r}_v$ of $ct$ in $\mathbf{P}_l$ / $\mathbf{P}_r$, and sends the membership witness to Bob / Alice. Then, Alice / Bob publishes $w^{ct, r}_v$ / $w^{ct, l}_v$ to $\mathrm{N}^P_l$ / $\mathrm{N}^P_r$. $\mathrm{N}^P_l$ / $\mathrm{N}^P_r$ verify the correctness of $w^{ct, r}_v$ / $w^{ct, l}_v$, and record it into $\mathbf{P}_l$ / $\mathbf{P}_r$. $\mathbf{P}_l$ system / $\mathbf{P}_r$ system will confirm that $ct$ has been recorded in $\mathbf{P}_r$ / $\mathbf{P}_l$. (see line 25-30 in Protocol \ref{Protocol_TransactionSynchronization})
\end{enumerate}

Fig. \ref{Figure_TransactionSynchronization} shows the process of UITS and JITS.

\textbf{Comparison}: We have a comparison between UITS and JITS. For the same points, UITS and JITS have the same result. Specifically, one cross-chain transaction $ct$ is recorded into two parachains $\mathbf{P}_l$ and $\mathbf{P}_r$, and $\mathbf{P}_l$ and $\mathbf{P}_r$ confirm the recording of $ct$ in each other.

For the different points, JITS does not need relay chain $\mathbf{R}$ to record $ct$, $w^{ct, l}$, and $w^{ct, r}$. For a cross-chain platform with numerous users, there will be a mass of $ct$, $w^{ct, l}$, and $w^{ct, r}$ generated per day, which may occupy the throughput of $\mathbf{R}$. If the users use JITS to have transaction synchronization, JITS will greatly reduce the throughput pressure of $\mathbf{R}$, which is more efficient than UITS.

However, compared with UITS, the prerequisite of JITS is more stringent. JITS requires Alice and Bob to cooperate to publish $ct$, $w^{ct, l}$ and $w^{ct, r}$. If one of Alice and Bob refuses to cooperate, $\mathbf{P}_l$ and $\mathbf{P}_r$ have to enable $\mathbf{R}$ again to accomplish transaction synchronization, in which JITS degrades to UITS. (see Section \ref{Subsection_Noncooperation})

\subsection{Cross-chain State Channel} \label{Subsection_Cross-chainStateChannel}

\begin{algorithm} [t]
	\footnotesize
	\setstretch{1.1}
	\caption{Channel Operation}
	\label{Protocol_ChannelOperation}
	
	\Fn{Opening}{
		// $U^l \leftarrow$ Alice ~~~// $U^r \leftarrow$ Bob\;
		\begin{tabular}{@{}p{1.4cm}@{}l@{}} \raggedright $U^l \Leftrightarrow U^r$ & : $\mathsf{Cosign} \rightarrow ct^{open}$\\  					    \end{tabular}\;
		\begin{tabular}{@{}p{1.4cm}@{}l@{}} \raggedright $U^l \Leftrightarrow U^r$ & : $\emph{JITS}(ct^{open})$\\                   					\end{tabular}\;
	}
	
	\vspace{0.5em}
	\Fn{Updating}{
		// $U^l \leftarrow$ Alice ~~~// $U^r \leftarrow$ Bob\;
		\begin{tabular}{@{}p{1.4cm}@{}l@{}} \raggedright $U^l \Leftrightarrow U^r$ & : $U^l$ and $U^r$ exchange signatures of $ct^{update}_m$\\			\end{tabular}\;
		\begin{tabular}{@{}p{1.4cm}@{}l@{}} \raggedright $U^l \Leftrightarrow U^r$ & : $U^l$ and $U^r$ exchange signatures of $pct^{update}_{m-1}$\\	\end{tabular}\;
	}
	
	\vspace{0.5em}
	\Fn{Closing}{
		\uIf{Joint closing}{
			// $U^l \leftarrow$ Alice ~~~// $U^r \leftarrow$ Bob\;
			\begin{tabular}{@{}p{1.4cm}@{}l@{}} \raggedright $U^l \Leftrightarrow U^r$ & : $\mathsf{Cosign} \rightarrow ct^{close}$\\					\end{tabular}\;
			\begin{tabular}{@{}p{1.4cm}@{}l@{}} \raggedright $U^l \Leftrightarrow U^r$ & : $\emph{JITS}(ct^{close})$\\              					\end{tabular}\;
		}
		\Else{
			// $U^i \leftarrow$ Alice or Bob\;
			\begin{tabular}{@{}p{0.5cm}@{}l@{}} \raggedright $U^i$ & : $\emph{UITS}(ct^{update}_m)$\\                                         			\end{tabular}\;
			\begin{tabular}{@{}p{0.5cm}@{}l@{}} \raggedright $U^i$ & : $U^i$ waits for $\Delta t$\\                                 					\end{tabular}\;
			\begin{tabular}{@{}p{0.5cm}@{}l@{}} \raggedright $U^i$ & : $\emph{UITS}(cct^{update}_m)$\\                                        			\end{tabular}\;
		}
	}
	
	\vspace{0.5em}
	\Fn{Disputing}{
		// $U^j \leftarrow$ Bob or Alice\;
		\Repeat{$ct^{update}_x$ $(0 < x < m)$ is recorded in $\mathbf{P}_j$}{
			\begin{tabular}{@{}p{0.5cm}@{}l@{}} \raggedright $U^j$ & : $U^j$ monitors $\mathbf{P}_j$\\                              					\end{tabular}\;
		}
		\begin{tabular}{@{}p{0.5cm}@{}l@{}} \raggedright $U^j$ & : $\emph{UITS}(pct^{update}_x)$\\                                           		    \end{tabular}\;
	}
\end{algorithm}

\subsubsection{Opening operation} By using JITS, Alice and Bob jointly publish an opening transaction $ct^{open}$ to $\mathbf{P}_l$ and $\mathbf{P}_r$ to achieve an opening operation. $ct^{open}$ will lock a part of Alice's and Bob's data in $\mathbf{P}_l$ and $\mathbf{P}_r$, and move the data into the cross-chain state channel as its initial state. The data can be tokens, assets, or private records of Alice and Bob. Then, the data can be processed in the channel without touching blockchains. Technically, it is also feasible to record $ct^{open}$ to $\mathbf{P}_l$ and $\mathbf{P}_r$ by using UITS. However, we suppose that Alice and Bob are cooperative at the beginning of the cross-chain state channel. Therefore, only the $ct^{open}$ published by JITS can open a cross-chain state channel. (see line 1-4 in Protocol \ref{Protocol_ChannelOperation})

\subsubsection{Updating operation} Updating operation includes two steps. Alice and Bob first draft an updating transaction $ct^{update}_m$, and exchange the signatures of $ct^{update}_m$. Second, Alice and Bob draft a punishment transaction $pct^{update}_{m-1}$ to invalidate $ct^{update}_{m-1}$, and exchange the signatures of $pct^{update}_{m-1}$ (see line 5-8 in Protocol \ref{Protocol_ChannelOperation}). The two steps are consistent with the existing intra-chain state channel \cite{Paper_LightningNetwork}. We do not have further illustrations.

\subsubsection{Closing operation} There are two categories of closing operations, which are joint closing and unilateral closing. In joint closing, Alice and Bob will publish closing transaction $ct^{close}$ to $\mathbf{P}_l$ and $\mathbf{P}_r$ by using JITS. According to the state in $ct^{close}$, Alice's and Bob's data in the channel will be returned back to their accounts in $\mathbf{P}_l$ and $\mathbf{P}_r$. Then, the cross-chain state channel is closed.

In unilateral closing, we assume that Bob attempts to unilaterally close the channel. He will publish the latest updating transaction $ct^{update}_m$ to $\mathbf{P}_l$ and $\mathbf{P}_r$ by using UITS. After a waiting time of $\Delta t$, Bob will publish the confirmation transaction $cct^{update}_m$ of $ct^{update}_m$ to $\mathbf{P}_l$ and $\mathbf{P}_r$ by using UITS. $cct^{update}_m$ will return the data of Alice and Bob in $ct^{update}_m$ back to their accounts in $\mathbf{P}_l$ and $\mathbf{P}_r$. Then, the cross-chain state channel is closed. (see line 9-18 in Protocol \ref{Protocol_ChannelOperation})

Noted, if Bob publish $cct^{update}_m$ within the waiting time of $\Delta t$, $cct^{update}_m$ is invalid. Besides, if Alice notices that Bob has published $ct^{update}_m$ by UITS, she can also publish $cct^{update}_m$ by using UITS, and $cct^{update}_m$ takes effect immediately without waiting a time of $\Delta t$.

\subsubsection{Disputing operation} In the process that Bob unilaterally closes the cross-chain state channel, there may be a malicious behavior that Bob publishes outdated transaction $ct^{update}_x$ $(0 < x < m)$ to $\mathbf{P}_l$ and $\mathbf{P}_r$, attempting to deny the latest transaction $ct^{update}_m$. For example, when $ct^{update}_x$ contains more tokens for Bob, compared with $ct^{update}_m$, Bob may choose to close the channel by $ct^{update}_x$ instead of $ct^{update}_m$, attempting to get more tokens back to his account. In this condition, a dispute occurs.

If Bob has the malicious behavior, $ct^{update}_x$ will be cross-chain transferred within a certain number of state synchronization rounds by UITS. Alice needs to remain online in $\mathbf{P}_l$ system to monitor whether a $ct^{update}_x$ is recorded into $\mathbf{P}_l$. If yes, Alice will have UITS to publish $pct^{update}_x$ to $\mathbf{P}_l$ and $\mathbf{P}_r$ within the time of $\Delta t$. $pct^{update}_x$ rejects $ct^{update}_x$ in $\mathbf{P}_l$ and $\mathbf{P}_r$, and closes the channel in the most favorable channel state for Alice, such as returning all the tokens in the channel to Alice's account, as a punishment to Bob. (see line 19-24 in Protocol \ref{Protocol_ChannelOperation})

\section{Security Analysis} \label{Section_SecurityAnalysis}
This section offers an informal security analysis to support the designs presented in Sections \ref{Section_Interpipe} and \ref{Section_BatchTransactionProof}. We will explore attack vectors, potential impacts, and ways to mitigate them.

\subsection{Hard Fork} \label{Subsection_HardFork}
A hard fork is a non-backward-compatible upgrade to a blockchain network that fundamentally changes its protocol, resulting in a split from the original chain and the creation of a new separate chain.

\subsubsection{Hard fork in parachain} When a hard fork occurs in left parachain $\mathbf{P}_l$ to create two separate parachains $\mathbf{P}_l^*$ and $\mathbf{P}_l^{**}$. Bob in the $\mathbf{P}_r$ system may attempt to establish two cross-chain state channels to $\mathbf{P}_l^*$ and $\mathbf{P}_l^{**}$ by only publishing one cross-chain transaction $ct$. Consequently, Bob's data in $\mathbf{P}_r$ can be used twice. (Alice may also have the same attempt when $\mathbf{P}_r$ has a hard fork.) To solve the problem, $\mathrm{N}^R$ need to select the main chain in $\mathbf{P}_l^*$ and $\mathbf{P}_l^{**}$. For example, it follows the chain with the most accumulated proof-of-work. Relay chain $\mathbf{R}$ only has cross-chain synchronization with the main chain. Subsequently, Bob can only establish a cross-chain state channel with Alice's account in one of $\mathbf{P}_l^*$ and $\mathbf{P}_l^{**}$.

\subsubsection{Hard fork in relay chain} There is another possibility that a hard fork occurs in relay chain $\mathbf{R}$ to create $\mathbf{R}^*$ and $\mathbf{R}^{**}$. The original cross-chain platform maintained by $\mathrm{N}^R$ is divided into two platforms respectively maintained by $\mathrm{N}^{*R}$ and $\mathrm{N}^{**R}$. Surprisingly, the relay chain hard fork has little impact on cross-chain state channel. Because relay chain is only an intermediary to transfer parachain state information, and it does not generate new information. If $\mathrm{N}^{*R}$ or $\mathrm{N}^{**R}$ have sufficiently large scale to ensure the security of $\mathbf{R}^*$ system or $\mathbf{R}^{**}$ system with $\alpha < 1/3$, the cross-chain synchronization can continue to support the operations in cross-chain state channel.

\subsection{Denial-of-service Attack} \label{Subsection_Denial-of-serviceAttack}
The existing cross-chain platform divides the relay node cluster $\mathrm{N}^R$ into $n$ relay node groups $\mathrm{N}^{R \sim}_i$. This division was intended to reduce communication and calculation overhead, as $\mathrm{N}^{R \sim}_i$ only needs to process information from one parachain. However, this design also reduces the number of nodes in each $\mathrm{N}^{R \sim}_i$, making it vulnerable to potential attacks. An adversary may attempt to take control of every node in $\mathrm{N}^{R \sim}_i$. The corrupted nodes may deny to pull the parachain state or generate state proof. Consequently, relay chain $\mathbf{R}$ cannot receive the real-time state information of a parachain $\mathbf{P}_i$, which in turn affects the other parachains, causing a synchronization interruption.

To address the denial-of-service attack, $\mathrm{N}^R$ adopt the random scheduling mechanism in XPull \cite{Paper_XPull} (see Section \ref{Subsection_Cross-chainStateTransfer}) to randomly select a new group $\mathrm{N}^{*R \sim}_i$ to replace $\mathrm{N}^{R \sim}_i$. There is a high possibility that $\mathrm{N}^{*R \sim}_i$ contains a portion of normal nodes to continue the cross-chain synchronization. To break the random scheduling mechanism, the adversary needs to control the majority of relay nodes $\mathrm{N}^R$. We assume it is hard to achieve. Moreover, if the adversary attempts to regain control of $\mathrm{N}^{*R \sim}_i$, another random scheduling can be executed to respond to it. We assume that the adversary is mildly adaptive \cite{Paper_Ouroboros}, i.e. the adversary cannot instantly corrupt every node in $\mathrm{N}^{*R \sim}_i$, and the corruption may only succeed after a sufficiently long period of time. Therefore, between each random scheduling, there is a time interval in which $\mathrm{N}^{*R \sim}_i$ includes at least one normal relay node to have cross-chain synchronization. Therefore, the interruption is only temporary, and the cross-chain synchronization can continue.

\subsection{Replay Attack} \label{Subsection_ReplayAttack}
Without adequate protection, a malicious party may attempt to replay cross-chain transactions. Specifically, the party publishes duplicate opening transaction $ct^{open}$ to open two cross-chain state channels without permission; the party publishes duplicate updating transaction $ct^{update}_m$ or closing transaction $ct^{close}$ to close the channel, attempting to return duplicate data to its account. A simple method to solve the problem is to add a unique identifier to each cross-chain transaction to prevent duplication.

\subsection{Counterfeiting} \label{Subsection_Counterfeiting}
A node in $\mathrm{N}^P_i$ may submit tampered parachain blocks to $\mathrm{N}^{R \sim}_i$, and a corrupted $\mathrm{N}^{R \sim}_i$ may generate counterfeit zero-knowledge proofs of the parachain state, attempting to synchronize the false state proofs into other blockchains. However, when $\mathbf{P}_i$ first established a cross-chain connection with relay chain $\mathbf{R}$, the parachain protocols and parachain node identities were made public to the cross-chain platform. Based on the tamper-proof property of blockchain, the tampered blocks or state proofs violate the parachain protocol or parachain node signature, which can be easily detected. Moreover, we assume the proportion of corrupted parachain nodes have $\alpha < 1/3$. Therefore, the adversary does not process enough hash rate (in PoW) or stake (in PoS) to create a replica of parachain, and consequently, the counterfeiting is hard to achieve.

\subsection{Eclipse Attack} \label{Subsection_EclipseAttack}
Bob may unilaterally close the cross-chain state channel by publishing an outdated updating transaction $ct^{update}_x$, attempting to deny the latest updating transaction $ct^{update}_m$ $(0 < x < m)$ (see Section \ref{Subsection_Cross-chainStateChannel}). At the same time, an adversary may create a fake network environment around Alice to prevent her from learning the publication of $ct^{update}_x$ through accessing $\mathbf{P}_l$, attempting to let Alice miss the opportunity to solve the dispute. For this problem, Alice needs to keep updating her local state of $\mathbf{P}_l$ by receiving new $\mathbf{P}_l$ blocks. If the new blocks cannot be received, she needs to find new P2P connections to ensure that at least one honest node of $\mathbf{P}_l$ can provide the service to prevent eclipse attacks.

\subsection{Conspiracy Attack} \label{Subsection_ConspiracyAttack}
Alice and Bob having a cross-chain state channel may attempt to collude to create blockchain data out of thin air. For example, between parachain $\mathbf{P}_l$ and $\mathbf{P}_r$, Alice and Bob each contribute 50 tokens to open a cross-chain state channel, noted $(50, 50)$, which includes 100 tokens in total. Next, Alice and Bob attempt to publish $ct^{update}_x$ to $\mathbf{P}_l$ to close the channel in a state of $(100, 0)$, by which Alice has 100 tokens returned to her account in $\mathbf{P}_l$. Then, Alice and Bob attempt to publish $ct^{update}_y$ to $\mathbf{P}_r$ to close the channel in a state of $(0, 100)$, by which Bob has 100 tokens returned to his account in $\mathbf{P}_r$. Finally, Alice and Bob have 200 tokens in total, with 100 tokens created out of thin air. The timeliness of cross-chain synchronization ensures that $\mathbf{P}_l$ and $\mathbf{P}_r$ learn the real-time states of each other. When a cross-chain transaction $ct$ is recorded in one of $\mathbf{P}_l$ and $\mathbf{P}_r$, $ct$ will be definitely recorded in the other one. Therefore, the $ct^{update}_x$ and $ct^{update}_y$ closing the same channel can be detected to conflict, and they will be refused by both $\mathbf{P}_l$ and $\mathbf{P}_r$, in which the conspiracy attack is hard to succeed.

\subsection{Noncooperation} \label{Subsection_Noncooperation}
The noncooperation may occur during JITS, in which one of Alice and Bob refuses to cooperate with the other one. For example, Alice may refuse to publish $ct$ to $\mathbf{P}_l$, send the membership witness $w^{ct, l}$ to Bob, nor publish $w^{ct, r}$ to $\mathbf{P}_l$. In these conditions, $\mathrm{N}^{R \sim}_l$ or $\mathrm{N}^{R \sim}_r$ will detect that $\mathbf{P}_l$ or $\mathbf{P}_r$ has only recorded $ct$ without recording $ct$ membership witness in the other parachain. If the problem persists for a long enough time, $\mathrm{N}^{R \sim}_l$ or $\mathrm{N}^{R \sim}_r$ will switch to UITS to have the synchronization of $ct$, using relay chain $\mathbf{R}$ to transfer $ct$ and its membership witnesses. Therefore, the noncooperation can only degrade JITS to UITS. The transaction synchronization can still be finished.

\section{Experiment and Evaluation} \label{Section_ExperimentAndEvaluation}
\subsection{Implementation} \label{Subsection_Implementation}
We develop a prototype implementation of Interpipe. The main process is written in Golang, Rust, and C++. In addition, the recursive proof algorithm utilizes Nova \cite{Paper_Nova}, and the hash algorithm adopts SHA-256. The blockchains, including a relay chain and two parachains, are based on Ouroboros consensus protocol \cite{Paper_Ouroboros}. We adjust the consensus slot to keep the block generation rate at 18 seconds per block. The system of relay chain $\mathbf{R}$ includes 120 relay nodes, denoted as $\mathrm{N}^R$. They are implemented with Intel Xeon Platinum 8280 @2.7GHz, 256GB DDR4 ECC DIMMs, and Windows Server operation system. In the initialization phase of relay chain system, two groups $\mathrm{N}^{R \sim}_l$ and $\mathrm{N}^{R \sim}_r$, each including 8 nodes, are randomly selected from $\mathrm{N}^R$ using a distributed randomness beacon based on Drand \cite{Paper_Drand}. The system of two parachains $\mathbf{P}_l$ and $\mathbf{P}_r$ respectively include 100 parachain nodes, denoted as $\mathrm{N}^P_l$ and $\mathrm{N}^P_r$. They are deployed on two hosts with Intel Core i9-13900K @3.0GHz, 64G DDRS 5200MHz XMP, and Windows operation system. Then, we deploy Protocol \ref{Protocol_StateSynchronization}, \ref{Protocol_TransactionSynchronization}, and \ref{Protocol_ChannelOperation} to Interpipe.

In each round of state synchronization, the state proofs of $\mathbf{P}_l$ and $\mathbf{P}_r$ are generated and transferred by relay nodes. The individuals within Interpipe will not have direct operations to state synchronization, as it automatically continues in the background to keep the consistency between parachains. We evaluate the proof generation efficiency in this process, and compare it with the existing work zkBridge \cite{Paper_zkbridge}, which is illustrated in Section \ref{Subsection_ProofGenerationEfficiency}. In the next process, two operators, respectively situated in $\mathbf{P}_l$ and $\mathbf{P}_r$ systems, first achieve the synchronization of a blank cross-chain transaction $ct$ by using UITS and JITS. Then, $ct$ is replaced by $ct^{open}$, $ct^{update}$, $ct^{close}$, and $pct^{update}$ to achieve the opening, updating, closing, and disputing operations to cross-chain state channel. We evaluate the throughput occupancy and time cost with different round duration, and compare the performance between Interpipe and the previous intra-chain state channel, which is illustrated in Section \ref{Subsection_ComparisonToIntra-chainStateChannel}.

\subsection{Proof Generation Efficiency} \label{Subsection_ProofGenerationEfficiency}
\begin{figure}[tbp]
	\centering
	\includegraphics[width=0.6\linewidth]{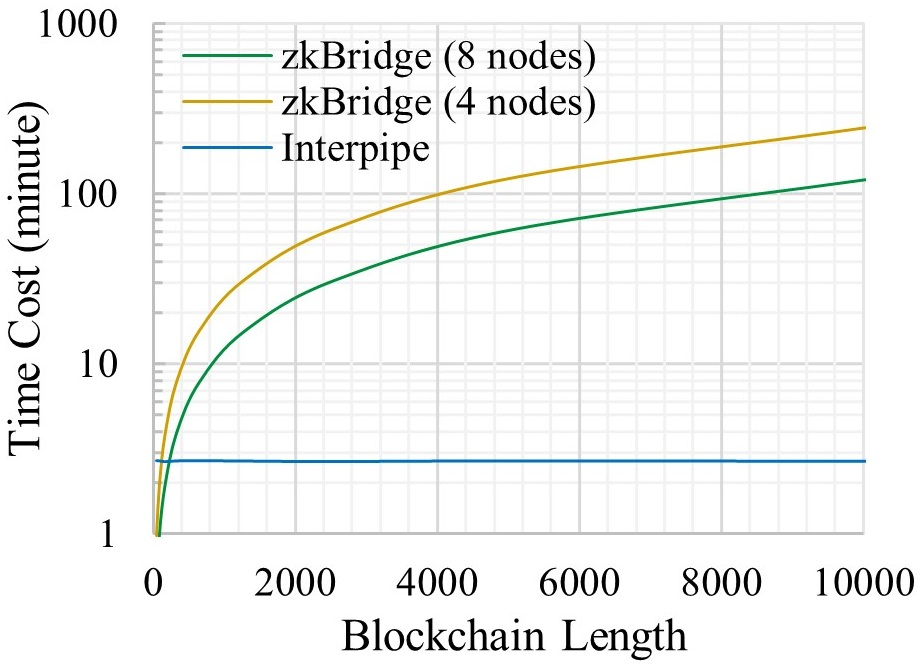}
	\caption{Proof generation time cost with different blockchain length}
	\label{Figure_ProofGenerationTimeCostWithDifferentBlockchainLength}
\end{figure}

We compare the proof generation time cost of Interpipe with the most recent work zkBridge \cite{Paper_zkbridge} by the blockchain length as a variable. Based on the strategy of zkBridge, we divide the arithmetic circuits in the parachain into $M$ copies, and distribute the $M$ copies to $M$ relay nodes for calculation, thereby increasing the proof generation speed by $M$ times. In our experiment (see Fig. \ref{Figure_ProofGenerationTimeCostWithDifferentBlockchainLength}), we set the value of $M$ to be 8 and 4, although this value can be larger in practical situations. However, zkBridge is only designed for one-round proof and does not make use of the proof generated in the previous round. Consequently, the arithmetic circuits in old blocks have to be recalculated in every round. It results in an increase in proving time as the blockchain length grows. In contrast, we use recursive SNARK to generate the state proof in Interpipe. To facilitate proof generation, the cross-chain transactions are included in a subtree of the Merkle tree in each parachain block. It takes about 7 seconds to finish the calculation of the arithmetic circuits in one block. The processes of proof generation and proof transfer can be carried out in parallel, with minimal impact on the time costs of each other. The experiment result shows that Interpipe's proving time remains relatively constant, as it does not require the recalculation of old blocks, and the number of new blocks generated in each round is almost constant.

\begin{figure}[tbp]
	\centering
	\includegraphics[width=0.6\linewidth]{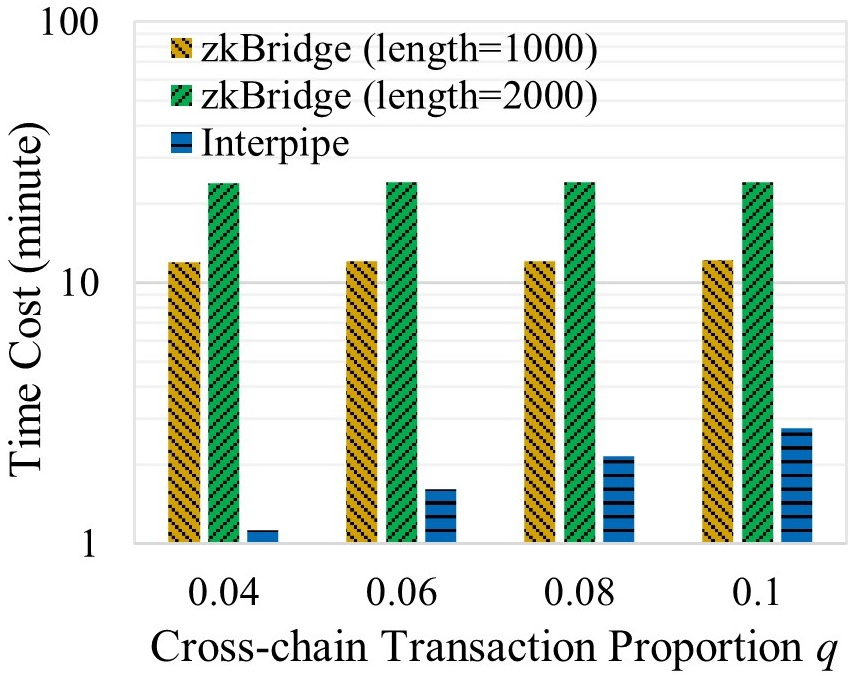}
	\caption{Proof generation time cost with different cross-chain transaction proportion}
	\label{Figure_ProofGenerationTimeCostWithDifferentCross-chainTransactionProportion}
\end{figure}

We also evaluate the proof generation time cost with different cross-chain transaction proportions $q$ (see Fig. \ref{Figure_ProofGenerationTimeCostWithDifferentCross-chainTransactionProportion}). For a parachain, each parachain block includes about 55 transactions, with a proportion of cross-chain transactions denoted as $q$ where $(0 < q < 1)$. In a practical situation, the value of $q$ depends on the preferences of all users in a parachain system. Considering most of the transactions in existing blockchains, such as Bitcoin and Ethereum, primarily focus on the internal affairs within the system, we set a relatively small value for $q$, in which $0 < q < 0.1$. The result shows that the time cost of Interpipe increases with $q$. Because batch transaction proof needs to prove every cross-chain transaction in the new block, leading to increased computation with a higher number of cross-chain transactions. In comparison, the time cost of zkBridge does not exhibit significant changes, as zkBridge is designed to prove individual cross-chain transactions. However, the time cost of Interpipe is still lower than the time cost of zkBridge, as the scale of cross-chain transactions in new blocks is much smaller than the scale of old blocks.

\subsection{Comparison to Intra-chain State Channel} \label{Subsection_ComparisonToIntra-chainStateChannel}
\begin{figure}[tbp]
	\centering
	\includegraphics[width=0.6\linewidth]{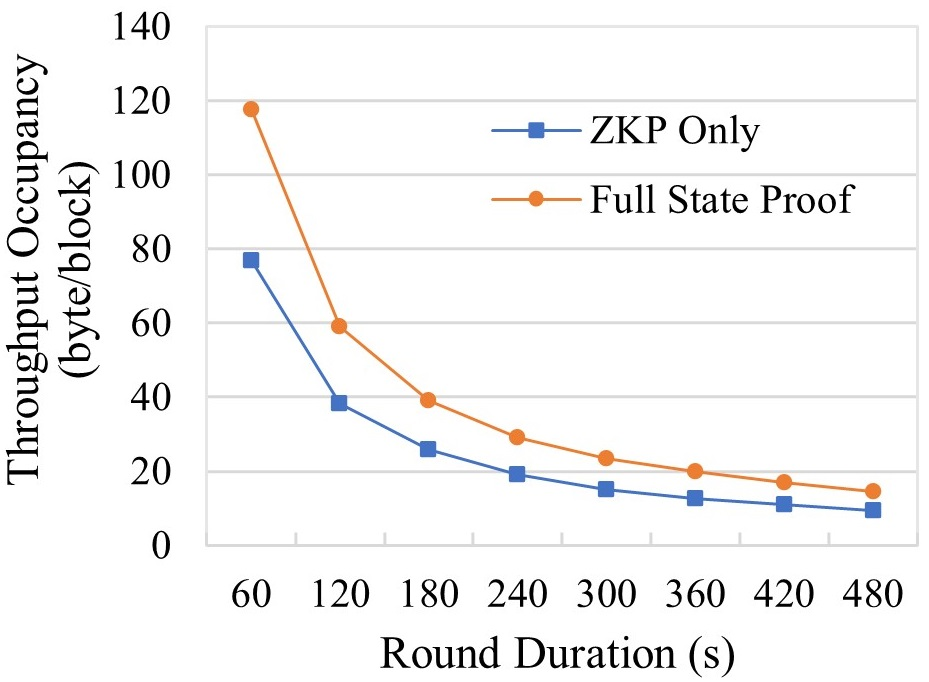}
	\caption{Throughput occupancy with different round duration}
	\label{Figure_ThroughputOccupancyWithDifferentRoundDuration}
\end{figure}

\begin{figure}[tbp]
	\centering
	\includegraphics[width=0.6\linewidth]{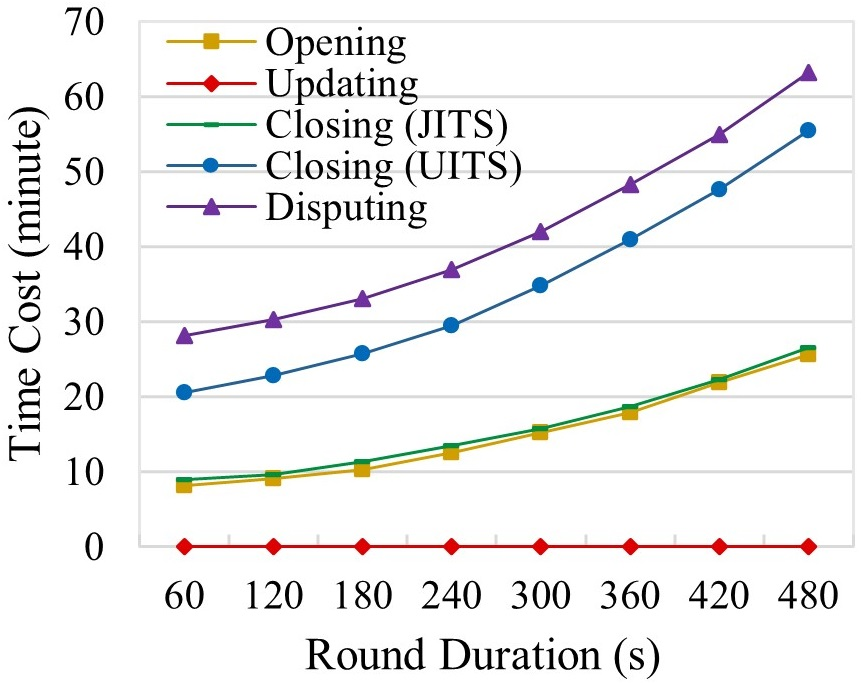}
	\caption{Operation time cost with different round duration}
	\label{Figure_OperationTimeCostWithDifferentRoundDuration}
\end{figure}

We begin by evaluating the performance of state synchronization. The duration of each round of state synchronization can be adjusted. Its value must be sufficiently large to prevent an excessively high frequency of state synchronization, which could lead to an accumulation of redundant state proofs in $\mathbf{R}$, thereby occupying its throughput (see Fig. \ref{Figure_ThroughputOccupancyWithDifferentRoundDuration}). Conversely, a longer round duration results in increased wait times for users (see Fig. \ref{Figure_OperationTimeCostWithDifferentRoundDuration}). The time costs associated with opening, closing, and disputing operations all rise with longer durations, while the time cost of updating operations remains constant, given that updates are executed off-chain. Consequently, there exists a trade-off between minimizing throughput occupancy and ensuring better service quality. The maintainers of a cross-chain platform need to find an appropriate balance in practical situations.

Adhering to our threat model, we maintain that a transaction within a blockchain achieves persistence when it reaches a depth of $k$ blocks, where $k = 9$. Concerning the trade-off in round duration, we have selected 240 seconds as the suitable duration for state synchronization. To compare the performance of the cross-chain state channel (CCSC) in Interpipe with the existing intra-chain state channel (ICSC) operating within a blockchain system, we deploy the ICSC protocol \cite{Paper_LightningNetwork} to a parachain system. The comparison of their performance is outlined in Table \ref{Table_ComparisonBetweenICSCAndCCSC}. The opening, closing, and disputing operations to CCSC indeed have a larger time cost than the same operations to ICSC. Because each operation to CCSC needs to have a transaction synchronization including 2 or more steps. A step refers to a cross-chain transaction or membership witness being recorded into a blockchain and becoming stable over time, with $Steps^\emph{JITS} = 2$ and $Steps^\emph{UITS} = 5$. Moreover, each transaction synchronization needs to wait for the state synchronization to enable cross-chain verification to the cross-chain transaction or membership witness, thereby incurring additional time costs. However, in updating operation, CCSC and ICSC exhibit similar time costs with relatively small values. As the updating operations play the main roles in off-chain interactions, if there are no urgent needs or malicious behaviors to close the channel, CCSC can be nearly as efficient as ICSC in most cases.

\begin{table}[tb]
	\centering
	\footnotesize
	\caption{Comparison between ICSC and CCSC}
	\label{Table_ComparisonBetweenICSCAndCCSC}
	\begin{tabular}{c|l|c|c}
		\toprule
										& \multicolumn{1}{c|}{\textbf{Operation}} & \textbf{Steps} & \textbf{Time Cost}         \\
		\midrule
		\multirow{5}{*}{\textbf{ICSC}}  & Opening                                 &      1         &    3.6 min				    \\
										& Updating     		                 	  &   	 0         &	62 ms				    \\ 
										& Closing (Joint)                      	  &   	 1         &	3.7 min				    \\
										& Closing (Unilateral)                 	  &  	 1         &	3.7 min                 \\
										& Disputing                            	  &  	 1         &	5.5 min				    \\
		\midrule
		\multirow{5}{*}{\textbf{CCSC}}  & Opening           	                  &      2         &    12.5 min			    \\
										& Updating     		                 	  &   	 0         &    105 ms			        \\ 
										& Closing (JITS)                       	  &   	 2         &    13.4 min			    \\
										& Closing (UITS)                       	  &   	 5         &    29.5 min                \\
										& Disputing                            	  &   	 5         &    36.9 min		   	    \\
		\bottomrule
	\end{tabular}
\end{table}

\section{Conclusion} \label{Section_Conclusion}
In this paper, we present a distributed cross-chain state channel scheme, called Interpipe. To meet the cross-chain verification needs of large-scale users, we propose a batch transaction proof scheme based on recursive SNARK. To achieve consistent operations between two blockchains, we propose a real-time cross-chain synchronization scheme. Based on the above designs, Interpipe offers protocols for opening, updating, closing, and disputing to cross-chain state channels. We have a security analysis of Interpipe, in which Interpipe can keep consistency, and withstand various existing attacks. The experimental results show that cross-chain state channels can be nearly as efficient as existing intra-chain state channels.

\bibliographystyle{IEEEtran}
\bibliography{References}

\section{Biography Section}
\vspace{-1.3cm}

\begin{IEEEbiography}[{\includegraphics[width=1in,height=1.25in,clip,keepaspectratio]{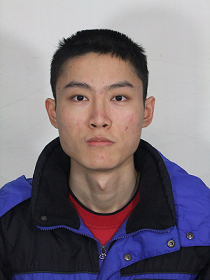}}]{Xinyu Liang}
	received the master's degree in Computer Science from Central China Normal University in 2018. He is currently working toward the Ph.D. degree in information security, Wuhan University, Wuhan. His research interests include blockchain and computer network.
\end{IEEEbiography}
\vspace{-1.3cm}

\begin{IEEEbiography}[{\includegraphics[width=1in,height=1.25in,clip,keepaspectratio]{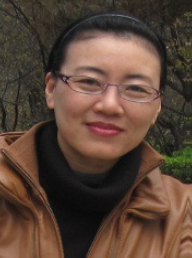}}]{Ruiying Du}
	received the BS, MS, PH. D degrees in computer science in 1987, 1994 and 2008, from Wuhan University, Wuhan, China. She is a professor at School of Cyber Science and Engineering, Wuhan University. Her research interests include network security, wireless network, cloud computing and mobile computing. She has published more than 80 research papers in many international journals and conferences, such as IEEE Transactions on Parallel and Distributed System, International Journal of Parallel and Distributed System, INFOCOM, SECON, TrustCom, NSS.
\end{IEEEbiography}
\vspace{-1.3cm}

\begin{IEEEbiography}[{\includegraphics[width=1in,height=1.25in,clip,keepaspectratio]{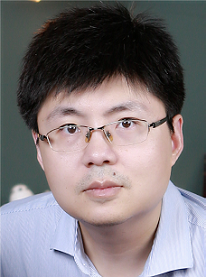}}]{Jing Chen}
	received the Ph.D. degree in computer science from Huazhong University of Science and Technology, Wuhan. He worked as a full professor in Wuhan University from 2015. His research interests in computer science are in the areas of network security, cloud security. He has published more than 100 research papers in many international journals and conferences, such as TDSC, TIFS, TMC, INFOCOM, TC, TPDS, et al. He acts as a reviewer for many journals and conferences, such as IEEE Transactions on Information Forensics, IEEE Transactions on Computers, IEEE/ACM Transactions on Networking.
\end{IEEEbiography}
\vspace{-1.3cm}

\begin{IEEEbiography}[{\includegraphics[width=1in,height=1.25in,clip,keepaspectratio]{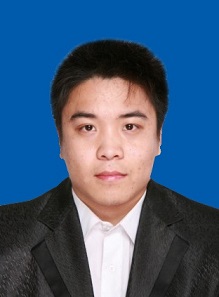}}]{Yu Zhang}
	was born in Shandong Province, China, in 1984. He received the Ph.D. degree from Wuhan University, China, in 2015. He is an engineer in Beijing Infosec Technologies Co., LTD. . His main research direction is network security.
\end{IEEEbiography}
\vspace{-1.3cm}

\begin{IEEEbiography}[{\includegraphics[width=1in,height=1.25in,clip,keepaspectratio]{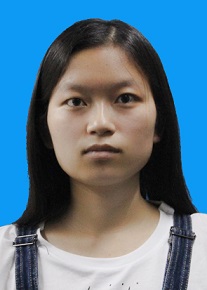}}]{Meng Jia}
	received the B.S. degree in computer science and technology from Wuhan University, Wuhan, China, in 2018. She is currently pursuing the Ph.D. degree with the School of Cyber Science and Engineering, Wuhan University, Wuhan, China. Her research interests include blockchain and applied cryptography.
\end{IEEEbiography}
\vspace{-1.3cm}

\begin{IEEEbiography}[{\includegraphics[width=1in,height=1.25in,clip,keepaspectratio]{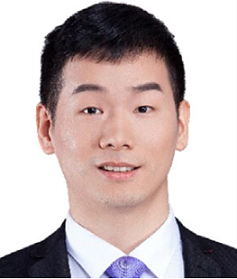}}]{Shixiong Yao}
	received the Ph.D. degree in information security in the School of Cyber Science and Engineering in Wuhan University. He is currently a lecturer with the school of computer, Central China Normal University, Wuhan. He has published papers in many international journals and conferences, such as the INFOCOM, TrustCom, the International Symposium on Emerging Information Security and Applications (EISA), the International Journal of Network Management (IJNM). His research interests are in the areas of Blockchain and Identity Management.
\end{IEEEbiography}
\vspace{-1.3cm}

\end{document}